\documentclass[twocolumn]{aastex63}
\usepackage{float,graphicx,amsmath,multirow}
\usepackage[version=3]{mhchem}
\usepackage{booktabs}
\usepackage{enumitem}

\newcommand\newtag[2]{#1\def\@currentlabel{#1}\label{#2}}

\received{March 8, 2021}
\revised{April 30, 2021}
\accepted{April 30, 2021}
\submitjournal{ApJL}

\shorttitle{Discovery of Indene with GOTHAM in TMC-1}
\shortauthors{Burkhardt et al.}
\graphicspath{{./}}
\begin{document}
%TC:ignore

\title{Discovery of the Pure Polycyclic Aromatic Hydrocarbon Indene ($c$-\ce{C9H8}) with GOTHAM Observations of TMC-1}
\author{Andrew M. Burkhardt}
\affiliation{Center for Astrophysics $\mid$ Harvard~\&~Smithsonian, Cambridge, MA 02138, USA}

\author{Kin Long Kelvin Lee}
\affiliation{Department of Chemistry, Massachusetts Institute of Technology, Cambridge, MA 02139, USA}
\affiliation{Center for Astrophysics $\mid$ Harvard~\&~Smithsonian, Cambridge, MA 02138, USA}

\author{P. Bryan Changala}
\affiliation{Center for Astrophysics $\mid$ Harvard~\&~Smithsonian, Cambridge, MA 02138, USA}

\author{Christopher N. Shingledecker}
\affiliation{Department of Physics and Astronomy, Benedictine College, Atchison, KS 66002, USA}

\author{Ilsa R. Cooke}
\affiliation{University of Rennes, CNRS, IPR (Institut de Physique de Rennes), UMR 6251, F-35000 Rennes, France}

\author{Ryan A. Loomis}
\affiliation{National Radio Astronomy Observatory, Charlottesville, VA 22903, USA}

\author{Hongji Wei}
\affiliation{Department of Physics and Astronomy, Benedictine College, Atchison, KS 66002, USA}

\author{Steven B. Charnley}
\affiliation{Astrochemistry Laboratory and the Goddard Center for Astrobiology, NASA Goddard Space Flight Center, Greenbelt, MD 20771, USA}

\author{Eric Herbst}
\affiliation{Department of Chemistry, University of Virginia, Charlottesville, VA 22904, USA}
\affiliation{Department of Astronomy, University of Virginia, Charlottesville, VA 22904, USA}

\author{Michael C. McCarthy}
\affiliation{Center for Astrophysics $\mid$ Harvard~\&~Smithsonian, Cambridge, MA 02138, USA}

\author{Brett A. McGuire}
\affiliation{Department of Chemistry, Massachusetts Institute of Technology, Cambridge, MA 02139, USA}
\affiliation{National Radio Astronomy Observatory, Charlottesville, VA 22903, USA}
\affiliation{Center for Astrophysics $\mid$ Harvard~\&~Smithsonian, Cambridge, MA 02138, USA}

\correspondingauthor{Andrew M. Burkhardt, Brett A. McGuire}
\email{andrew.burkhardt@cfa.harvard.edu, brettmc@mit.edu}

\begin{abstract}

Polycyclic Aromatic Hydrocarbons (PAHs) have long been invoked in the study of interstellar and protostellar sources, but the unambiguous identification of any individual PAH has proven elusive until very recently. As a result, the formation mechanisms for this important class of molecules remain poorly constrained. Here we report the first interstellar detection of a pure hydrocarbon PAH, indene (\ce{C9H8}), as part of the GBT Observations of TMC\nobreakdash-1: Hunting for Aromatic Molecules (GOTHAM) survey. This detection provides a new avenue for chemical inquiry, complementing the existing detections of CN-functionalized aromatic molecules. From fitting the GOTHAM observations, indene is found to be the most abundant organic ring detected in TMC\nobreakdash-1 to date. And from astrochemical modeling with \texttt{nautilus}, the observed abundance is greater than the model's prediction by several orders of magnitude suggesting that current formation pathways in astrochemical models are incomplete. The detection of indene in relatively high abundance implies related species such as cyanoindene, cyclopentadiene, toluene, and styrene may be detectable in dark clouds.

%PAHs. Where are they and where do they come from? Well we found a bunch of almond-smelling ones previously. But what if you're allergic? Here, try our new 100\% pure organic hydrocarbon PAHs. Brought to you by your local friendly GOTHAM farmers: reaping new detections for you since 2018. \textit{We aren't the heroes you want, but the ones you deserve.}

\end{abstract}
\keywords{Astrochemistry, Dark interstellar clouds, Interstellar molecules, Polycyclic aromatic hydrocarbons}

%TC:endignore

\section{Introduction}
\label{sec:intro}

Aromaticity and organic rings are vital components of both interstellar and terrestrial chemistry. Because aromatic compounds exhibit exceptional stability, many play key roles in the chemistry of biological systems, including the single- or double-ringed heterocyclic primary nucleobases. In the interstellar medium (ISM), 10--25\% of all carbon may be locked in large multi-ring hydrocarbons called Polycyclic Aromatic Hydrocarbons (PAHs) based on the strength and ubiquity of the Unidentified Infrared Bands (UIRs) \citep{Tielens:2008}. PAHs are believed to coalesce into interstellar dust grains and the carbonaceous feedstock for planetary bodies. While aromaticity is a key structural motif on virtually all size scales from protoplanetary disks \citep{Dullemond:2007,Perez-Becker:2011} to the largest galaxies \citep{Smith:2007}, their formation has remained poorly constrained for decades due to the inability to detect any specific PAHs.

Two contrasting pathways have been proposed for PAH production. In the `top-down' scenario, large, multi-ringed species like fullerenes, PAHs, and Mixed Aromatic-aliphatic Organic Nanoparticles are formed in the hot, dense outflows of carbon-rich evolved stars \citep{Tielens:2008,Kwok:2011iv,Martinez:2019ud}.  Processing of these species in diffuse clouds subsequently then generates smaller, simpler species in high abundances, although this processing may favor the destruction of smaller rings \citep{montillaud_evolution_2013,chabot_coulomb_220}. This highlights the potential importance of a `bottom-up' scenario by which complex molecules are built up from small carbon-chains in the cold, shielded environments of dark clouds. 

Abundance measurements of individual aromatic molecules in tandem with astrochemical modeling may be crucial to differentiate between these two scenarios. The UIRs, however, are an aggregate of nearly degenerate vibrational features produced from the \ce{C-C} and \ce{C-H} bonds of the full range of PAHs in a given source, thus precluding any unique identifications \cite{Tielens:2008,Bauschlicher:2018ej}. A single infrared vibrational transition of the simplest aromatic molecule, benzene ($c$-\ce{C6H6}), has been detected towards a handful of evolved sources \citep{Cernicharo:2001,Cami:2010fi,Berne:2013ow}, but has limited utility as atmospheric absorption prevents its ground-based observability. In contrast, a molecule's pure rotational spectroscopy at radio and millimeter frequencies provides detailed information into its excitation conditions and abundance \citep{McGuire:2018mc}. However, the high symmetry and lack of permanent electric dipole moment of key aromatic building blocks like benzene and naphthalene ($c$-\ce{C10H8}) prevent their radio and millimeter detections. As such, direct detection of small polar aromatic molecules could greatly advance our understanding of PAH formation.

The GBT Observations of TMC\nobreakdash-1: Hunting for Aromatic Molecules (GOTHAM) survey \citep{McGuire:2020bb}, a high-sensitivity spectral line study spanning approximately 2--12 and 18--34\,GHz using the 100\,m Robert C. Byrd  Green Bank Telescope (GBT), has sought to critically examine the presence of aromatic molecules toward the prototypical dark cloud Taurus Molecular Cloud 1 (TMC\nobreakdash-1) Cyanopolyyne Peak (CP). TMC\nobreakdash-1 CP is an ideal source to detect complex carbon molecules as it is a cold ($T$\,$\sim$\,10\,K), quiescent ($\Delta V$\,$\sim$\,0.12\,km\,s$^{-1}$) cloud that is chemically-rich ($\sim$50 new molecular detections) and less line-dense than other prototypical astrochemical sources (e.g. IRC+10216, Orion KL, Sgr B2N, IRAS16293). Over the past several years, surveys of TMC-1 have resulted in many new molecular detections \citep{McGuire:2017,Burkhardt:2018ka,Xue:2020aa, McGuire:2020bb, Loomis:2021aa, Lee:2021aa, Shingledecker:2021,Marcelino:2021,Cernicharo:2021}, including several functionalized one and two-ringed species: benzonitrile ($c$-\ce{C6H5CN}, \citealt{McGuire:2018it}), cyanocyclopentadiene (1-\ce{C5H5CN}, \citealt{McCarthy:2020aa}; 2-\ce{C5H5CN}, \citealt{Lee:2021bb}), and cyanonaphthalene (1- and 2-\ce{C10H7CN}, \citealt{McGuire:2021aa}). Other ongoing surveys have led to the detection of additional  carbon chains in TMC\nobreakdash-1, further expanding the chemical inventory of this remarkably rich source  \citep{Marcelino:2021,Cernicharo:2021}. As part of A Rigorous K/Ka-Band Hunt for Aromatic Molecules (ARKHAM) survey, it has also been found that benzonitrile is ubiquitous among prestellar sources beyond TMC\nobreakdash-1 \citep{Burkhardt:2021}. These many new detections of aromatic molecules and their carbon-chain precursors have opened up new avenues to study this previously unconstrained chemical regime.

Critical in enabling these detections, the presence of a cyano (--CN) group dramatically increases a molecule's electric dipole moment relative to its pure hydrocarbon counterpart. Based on extensive laboratory and modeling efforts, CN-addition to ringed molecules is believed to be barrierless and highly efficient under interstellar conditions \citep{parker_low_2012,Lee:2019dh,Cooke:2020we}. Thus, functionalized rings could be used as proxies for symmetric hydrocarbon aromatics like benzene and naphthalene. However, there has not been a robust way to test this theory observationally due to the lack of pure hydrocarbon PAH detections. Given the known presence of individual five- and and six-membered rings, a polar molecule containing both could exist especially when larger PAHs (e.g. cyanonaphthalene) are also abundant. Indene (\ce{C9H8}), composed of both a five- and six-membered ring with a permanent dipole moment of $\sim$0.73\,D (see Tab.~\ref{tab:indene_spectroscopy}), is a possible target for radio observations. Recent laboratory and theoretical work has suggested that this molecule may be formed efficiently enough to be detected by radio observatories toward dark clouds  \citep{Doddipatla:2021}. Due to it's reasonable dipole moment, the detection of indene and cyanoindene could be used to infer a ratio between pure hydrocarbon aromatics and their cyano-derivatives. Among pure hydrocarbons, closed rings have additional low frequency vibrational modes that can be used to more efficiently dissipate energy from dissociative photons compared to rings functionalized by atomics species or carbon chains. In particular, indene presents perhaps the simplest closed ring molecule larger than benzene that both is energetically stable and has a non-zero dipole moment. Here, we present the first interstellar detection of indene, the first purely hydrocarbon PAH detected in space, using GOTHAM survey data and improved laboratory spectroscopic measurements. Preliminary gas-grain chemical modeling is then used to provide an initial interpretation of the results. 

\section{Observations}
\label{sec:obs}
These observations and subsequent analysis strategy are discussed in detail in \citep{McGuire:2020bb} and \citep{McGuire:2021aa}, Here, we provide a brief summary.  Observations were performed with the 100-m  Robert C. Byrd  Green Bank Telescope with the project codes GBT17A\nobreakdash-164,  GBT17A\nobreakdash-434, GBT18A\nobreakdash-333, GBT18B\nobreakdash-007, and data from project GBT19A\nobreakdash-047 acquired through June 2020. The pointing was centered on TMC\nobreakdash-1 CP at (J2000) $\alpha$~=~04$^h$41$^m$42.50$^s$ $\delta$~=~+25$^{\circ}$41$^{\prime}$26.8$^{\prime\prime}$. The spectra were obtained through position-switching to an emission-free position 1$^{\circ}$ away. Pointing and focusing was refined every 1--2 hours on the calibrator J0530+1331. Flux calibration was performed with an internal noise diode and Very Large Array (VLA) observations (Project TCAL0003) of the same calibrator used for pointing, resulting in a flux uncertainty of ${\sim}20$\% \citep{McGuire:2020bb}. In total, the spectral coverage of the GOTHAM DR2 survey spans 2--12 and 18--34\,GHz, with an RMS noise level of 2--20\,mK when convolved to a uniform velocity resolution of 0.05\,km\,s$^{-1}$. 

\section{Laboratory Spectroscopy}
\label{sec:spec}

Prior laboratory measurements of indene rest frequencies~\citep{Li1979} include only 12 rotational transitions in the 21--26\,GHz range with a relatively large uncertainty of 0.1\,MHz, which is insufficiently precise for generating a catalog of narrow-line transitions in TMC\nobreakdash-1 ($\Delta V$\,$\sim$10\,kHz at 30\,GHz). Using a cavity-enhanced Fourier transform microwave spectrometer co-axially coupled to a supersonic jet expansion of $0.1$\% indene seeded in neon~\citep{Grabow:2005kf}, we remeasured and substantially extended the data set of indene rotational transitions from 5 to 40\,GHz with a uniform frequency accuracy of 2\,kHz---a 50-fold improvement over previous measurements. In total, 189 $a$- and $b$-type transitions with rotational quantum numbers $J \leq 13$ and $K_a \leq 6$ were observed and fitted to the Watson A-reduced Hamiltonian including quartic centrifugal distortion terms~\citep{Watson:1977vk}. The optimized rotational constants, which resulted in a normalized fit RMS error of $0.43$, are summarized in Table~\ref{tab:indene_spectroscopy}.

\begin{table}[t]
    \centering
    \caption{Spectroscopic constants of indene. The updated laboratory rest frequencies were fitted to Watson's A-reduced (I$^r$) quartic centrifugal distortion Hamiltonian. All values are in MHz unless noted. Values in parentheses denote one standard deviation and apply to the last digits of the constants. Dipole moments are calculated at the $\omega$B97X-D/6-31+G(d) level of theory. Dipole moments have a nominal Bayesian uncertainty of +/-0.5 D \cite{Lee:2020}.}
    \label{tab:indene_spectroscopy}
    \begin{tabular}{r@{}lr@{.}lr@{.}l}
        \toprule
    \multicolumn{2}{c}{Parameter} & \multicolumn{2}{c}{This work} & \multicolumn{2}{c}{\cite{Li1979}} \\
    \midrule
    \multicolumn{2}{c}{$A$} & 3775&03769(11) & 3775&01(13)\\
    \multicolumn{2}{c}{$B$} & 1580&864105(40)& 1580&86(1)\\
    \multicolumn{2}{c}{$C$} & 1122&248364(31)& 1122&24(1)\\
    $\Delta_J$&$ \times 10^6$ & 33&83(12)& \multicolumn{2}{c}{--} \\
    $\Delta_{JK}$&$ \times 10^6$ & 47&53(82) & \multicolumn{2}{c}{--}\\
    $\Delta_{K}$&$\times 10^6$ & 257&5(44)& \multicolumn{2}{c}{--}\\
    $\delta_J$&$ \times 10^6$ & 10&215(69)& \multicolumn{2}{c}{--}\\
    $\delta_K $&$\times 10^6$ & 86&9(14)& \multicolumn{2}{c}{--}\\
    \multicolumn{2}{c}{$N_{\rm meas}^a$} & \multicolumn{2}{c}{189} & \multicolumn{2}{c}{12}\\
    \multicolumn{2}{c}{$\mu_a$} & 0&59\,D & \multicolumn{2}{c}{--}\\
    \multicolumn{2}{c}{$\mu_b$} & 0&43\,D & \multicolumn{2}{c}{--}\\
    \bottomrule 
    \multicolumn{6}{l}{$^a$ The number of measurements included in the fit} \\
    \end{tabular}
\end{table}

\section{Analysis and Results}
\label{sec:results}

\subsection{Observational Results, Velocity Stacking, and Matched Filtering}
\label{sec:stacking}
As of the second data release of GOTHAM, no individual transitions of indene were bright enough for identification. To robustly detect molecules with many faint transitions within the GOTHAM coverage, we utilize the spectral stacking and matched filtering procedure detailed in \citet{Loomis:2021aa}. In summary, a small spectral window is extracted for all indene transitions in the GOTHAM spectral coverage that are predicted to be at least 0.1\% of the strongest transition in the catalog, provided that there is no interloping emission ($>$3.5$\sigma$) present in the spectrum. In this case, a total of 2048 rotational transitions of indene met these criteria with two interloping transitions detected. The windows are subsequently combined in velocity space, each weighted by the observational RMS and the predicted flux.

We additionally carry out a forward modeling procedure using \texttt{molsim} \citep{lee_molsim_2020} which simulates the molecular emission with a set of model parameters including the source size (SS)%---used for estimating beam dilution effects---
, radial velocity ($v_{lsr}$), column density ($N_T$), excitation temperature ($T_{ex}$), and the line width ($\Delta $V). In accordance with previous high resolution observations of TMC\nobreakdash-1, our model also includes contributions from four distinct velocity components, which corresponds to independent treatments of SS, $v_{lsr}$, and $N_{T}$, resulting in a total of 14 modeling parameters. To properly account for uncertainty and covariance between model parameters, we use affine-invariant Markov Chain Monte Carlo (MCMC) sampling as implemented in \texttt{emcee} \citep{foreman-mackey_emcee_2013}, estimating the likelihood for each given set of model parameters. As with previous work on aromatic species \citep{McCarthy:2020aa, Lee:2021bb,McGuire:2021aa}, the posterior for benzonitrile reported in \citet{McGuire:2021aa} was used as the prior distribution for the indene MCMC modeling. This modeling choice explicitly makes the assumption---on grounds of chemical similarity---that indene shares similar physico-chemical modeling parameters with benzonitrile. While certain aspects may not necessarily transfer, for example the appropriate number of velocity components, the lack of individual line detections for indene as well as the lack of interferometric data constricts the avenues available for analysis and the interpretability of the model parameters.

\begin{figure*}[bt]
    \centering
    \includegraphics[width=\textwidth]{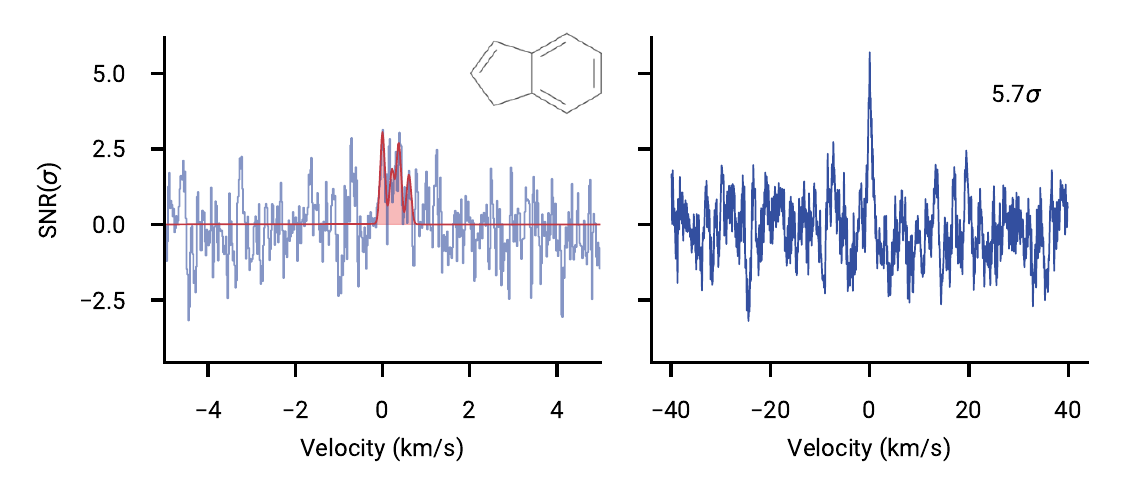}
    \caption{Velocity stacked and matched filter spectra of indene (pictured). The intensity scales are the signal-to-noise ratios (SNR) of the response functions when centered at a given velocity. The ``zero'' velocity corresponds to the channel containing the highest intensity within the simulated spectra to account for blended spectroscopic transitions and variations in velocity component source sizes. (\emph{Left}) The stacked spectra from the GOTHAM DR2 data are display in blue, overlaid with the expected line profile in red from our MCMC fit to the data.  The signal-to-noise ratio is on a per-channel basis. (\emph{Right}) Matched filter response obtained from cross-correlating the velocity stacks in the left panel; value annotated corresponds to the peak impulse response of the matched filter.}
    \label{fig:indene_stack}
\end{figure*}

Following convergence of the sampling, the resulting posterior was analyzed using the \texttt{arviz} suite of routines \citep{kumar_arviz_2019}, and the expected emission from indene is simulated using the posterior mean parameters shown in Table \ref{tab:indene_fits}. The left of Figure \ref{fig:indene_stack} overlays the simulated and observed velocity stacks, with a peak SNR of ${\sim}3\sigma$. To quantitatively assess the likelihood of detection, the observational velocity stack is cross-correlated with the simulated emission as a form of matched filter process, shown on the right panel of Figure \ref{fig:indene_stack} where the peak impulse response ($\sigma$) corresponds to the statistical significance of the detection. As with previous GOTHAM analyses, we adopt a $5\sigma$ threshold to classify a firm detection \citep{Loomis:2021aa}; here, the matched filter produces a $5.7\sigma$, signifying a firm detection. As some indene transitions have closely spaced lines, such as from dense sets of K-ladder transitions, we treat groups of transitions as a single feature for the purposes of stacking. This results in the potential loss of the interpretability of velocity structure derived from the stacked spectrum line profile. However, since the significance of the detection is determined by the matched filter response, this does not hinder the analysis, as the same stacking procedure is applied to the observed data and simulated spectra, which is then used to determine the stack weighting and matched filter profile. 

Regarding the MCMC derived parameters, Table \ref{tab:indene_fits} shows that the radial velocities, modeled source size, excitation temperature, and line width are consistent with similar molecules (that are not optically thick) detected in TMC-1, such as cyanocyclopentadiene \citep{Lee:2021bb,McCarthy:2020aa} and cyanonaphthalene \citep{McGuire:2021aa}---for example, $T_{\text{ex}}$\,$\sim$8--9\,K and $\Delta V$\,$\sim$0.12\,km\,s$^{-1}$. The posterior radial velocities are largely similar to that of benzonitrile for all except the 5.421\,km\,s$^{-1}$ velocity component, diverging over a line width's worth from the prior value (5.57\,km\,s$^{-1}$ from \citet{McGuire:2021aa}). From this it appears that the number of velocity components may need to be reassessed, either through observational constraints (as mentioned previously) or through Bayesian model selection. For the former, we are currently undertaking observations with the VLA, which we expect will provide more insight into the resolvable velocity structure embedded in TMC\nobreakdash-1, and, for the latter, should provide physically interpretable MCMC modeling beyond nuisance parameters.

\begin{table}[bt]
    \centering
        \caption{Summary statistics of the marginalized indene posterior. The quoted uncertainties correspond to the 95\% highest posterior density. The total column density is derived from combining the column densities of each component.}
    \begin{tabular}{ c c c c c}
    \toprule
    	$v_{lsr}$	&	Size	&	$N_T$	&	$T_{ex}$	&	$\Delta V$		\\
    	(km\,s$^{-1}$)&	($^{\prime\prime}$)	&	(10$^{12}$cm$^{-2}$) & (K)	&  (km\,s$^{-1}$)\\ 
    	\midrule
         $5.421^{+0.040}_{-0.040}$  &   $154^{+205}_{-146}$ &   $1.84^{+2.92}_{-1.48}$   &  \multirow{4}{*}{$8.55^{+0.22}_{-0.23}$}    & \multirow{4}{*}{$0.123^{+0.016}_{-0.016}$}  \\
         $5.654^{+0.024}_{-0.023}$  &   $199^{+166}_{-177}$  &   $2.94^{+1.00}_{-1.01}$   &  &   \\
         $5.811^{+0.030}_{-0.031}$  &   $255^{+145}_{-157}$  &   $1.95^{+0.68}_{-0.67}$   &      &   \\
         $6.027^{+0.014}_{-0.013}$   &   $252^{+148}_{-150}$  &   $3.24^{+0.70}_{-0.67}$   &      &   \\
    \midrule
    \multicolumn{5}{c}{$N_T$ (Total): $9.60^{+4.33}_{-1.57}\times10^{12}$\,cm$^{-2}$} \\
    \bottomrule
    \end{tabular}
    \label{tab:indene_fits}
\end{table}

\subsection{Astrochemical Modeling}
\label{sec:model}

To study the formation of indene and its related species, we adapted the three-phase chemical network model \texttt{nautilus} v1.1 code \citep{Ruaud:2016}. Originally based on the \textsc{KIDA} network, the extended aromatic and carbon-chain network has been expanded to include reactions related to this and other species detected using GOTHAM data. In all these works, our model was able to very successfully reproduce the observed abundances of the new carbon-chain molecules \citep{Xue:2020aa, McGuire:2020bb, Loomis:2021aa, Shingledecker:2021}, while systematically underproducing the abundance of cyclic molecules \citep{McCarthy:2020aa, Burkhardt:2021, McGuire:2021aa}. 
 
The physical conditions of the model are consistent with the previous modeling work of TMC\nobreakdash-1 as part of the GOTHAM survey, originally constrained by \citep{hincelin_oxygen_2011}, with a gas and grain temperature of $T_{\text{gas}}$=$T_{\text{grain}}$=10\,K, a gas density of $n_{\text{H}_2}$=2$\times$10$^4$\,cm$^{-3}$, a visual extinction of $A_V$=10, and a cosmic ray ionization rate of $\zeta_{\text{CR}}$=1.3$\times$10$^{-17}$\,s$^{-1}$. The initial elemental abundances were consistent with previous versions of this model that were constrained in \citet{Loomis:2021aa}.%, including the recently re-determined C/O ratio based on fitting the peak abundances of the cyanopolyyne family to the observed column densities reported in the GOTHAM second data release (DR2) in \citet{Shingledecker:2021}. 

Building upon the previous networks used as part of the GOTHAM survey (most recently \citet{Shingledecker:2021}), we included additional reactions to account for the production and destruction of indene and cyanoindene  ($c$-\ce{C9H7CN}). In particular, we incorporated the formation pathways of indene known as the methylidyne addition-cyclization-aromatization (MACA) mechanism discussed in \citet{Doddipatla:2021}. Specifically, we added the gas-phase production of indene through the successive bi-molecular barrierless reactions from toluene (\ce{C6H5CH3}) to styrene (\ce{C6H5C2H3}), which are in turn produced by semi-saturated chains, including 1,3-pentadiene (\ce{C5H8}), 3-hexen-1-yne	(\ce{C6H8}), 1-propynyl (\ce{CCCH3}), and 1,3-butadiene (\ce{CH2CHCHCH2}) (See Table~\ref{tab:rxs} Reactions~1--4%\ref{rx:d21prodi}--\ref{rx:d21prodf}
). While several of these species were previously studied in this network \citep{Burkhardt:2021,Shingledecker:2021}, we have included additional processes for the production of \ce{C5H8} and \ce{C6H8} and their precursors \ce{C3H7}, \ce{C3H8}, and \ce{C4H8} using pathways and rates measured or computed in the literature \citep[and references therein]{Loison:2017,Hebrard:2009,Morales:2010,anicich_index_2003} (See Table~\ref{tab:rxs} Reactions~12--83%\ref{rx:kidai}--\ref{rx:kidaf}
). Also from \citet{Doddipatla:2021}, we added the destruction of indene through the subsequent production of a limited subset of larger PAHs, including naphthalene ($c$-\ce{C10H8}) and cyanoindene, using the average reaction rate value reported (See Table~\ref{tab:rxs} Reactions~5--11%\ref{rx:d21desti}--\ref{rx:d21destf}
).%
 For each of the bi-molecular molecular reactions described above, corresponding ice-surface and mantle reactions were also included in the network, similar to \citet{Doddipatla:2021}.

Finally, we have estimated the ion-neutral destruction of indene, cyanoindene, \ce{C4H8}, \ce{C5H8}, and \ce{C6H8} using the Langevin formula \citep{woon_quantum_2009} (See Table~\ref{tab:rxs} Reactions~84--178%\ref{rx:thisi}--\ref{rx:thisf}
). For indene and cyanoindene, the predicted pathways are based on related processes for naphthalene and cyanonaphthalene \citep{McGuire:2021aa}, excluding reactions with \ce{H3+}, \ce{H3O+}, and \ce{HCO+} as they mostly produce protonated forms that primarily dissociate back into the original species \citep{Milligan:2002}. The products for the dissociation of \ce{C4H8}, \ce{C5H8}, and \ce{C6H8} are based on the similar pathways for \ce{C3H8}.  The branching ratios are assumed to be equal among the proposed products. While these estimated production channels may not be fully representative of the true branching ratios, they do provide an estimate for the destruction of these species to test the feasibility of the model to reproduce the observed abundance of indene. %These new reactions and their rates are summarized in Table~\ref{tab:rxs}.% and the polarizability and dipole moments used in the calculations are provided in Table~\ref{tab:dipole_n_polar}.
 
\begin{figure}[tb]
    \centering
    \includegraphics[width=0.5\textwidth]{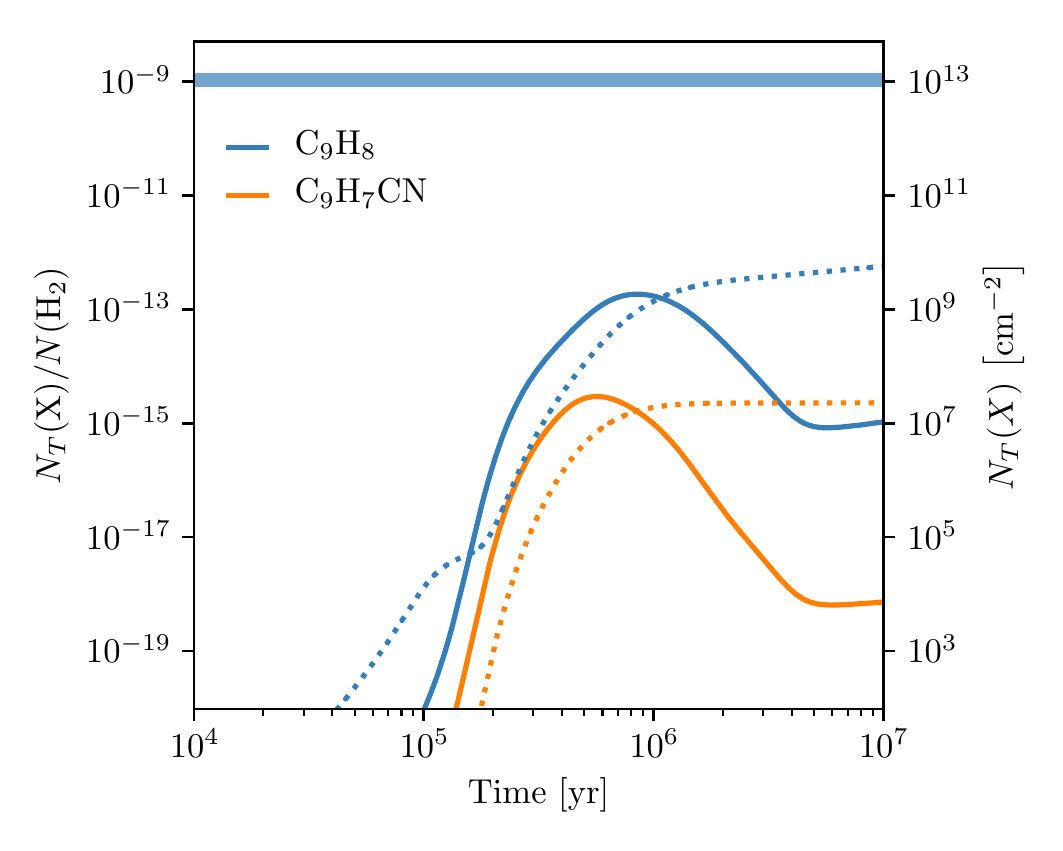}
    \caption{Simulated abundance and column density of indene (blue) and cyanoindene (orange) in the from \texttt{nautilus} chemical models in the gas phase (solid) and grain surface (dotted) in comparison to the observed indene column density with uncertainties as a blue horizontal bar.}
    \label{fig:indene_model}
\end{figure}

The results of this model can be seen in Figure~\ref{fig:indene_model} with the simulated abundance of indene compared to the observed abundance toward TMC-1 assuming a $N_{\text{H}_2}$\,$\sim$10$^{22}$\,cm$^{-2}$, showing that this gas-phase production route is insufficient to reproduce the observed value from GOTHAM by several orders of magnitude. This overabundance of observed molecules is consistent for all detected cyclic molecules thus far toward TMC\nobreakdash-1, as discussed in detail by \citet{Burkhardt:2021}. The chemical implications and potential improvements to this model are discussed in Section~\ref{sec:implications}.

\section{Discussion}
\label{sec:disc}

\subsection{Assay of Detected Interstellar Ringed Species }
\label{sec:assay}

With the discovery of an abundant pure PAH here, we can begin to make more quantitative comparisons amongst the six ring molecules in TMC\nobreakdash-1 to date. Their updated derived column densities from the GOTHAM DR2 survey are summarized in Table~\ref{tab:aromatic cols}, showing indene is the most abundant aromatic molecule discovered in TMC\nobreakdash-1 by a factor of 5--50. This suggests the addition of a cyano-group limits the abundance of this family relative to their pure hydrocarbon counterparts. Significant laboratory and theoretical effort has shown that the reactions between ringed molecules and CN are thought to be barrierless and occur on virtually every collision \citep{parker_low_2012,Lee:2019dh,Cooke:2020we}. As such, these cyclic species with CN groups may be crucial proxies for their pure hydrocarbon precursors. The detection of the indene is a vital step to calibrate this cyano/pure ring relation. 

Furthermore, while a larger abundance of pure PAHs relative to functionalized PAHs appears intuitive, the same cannot be said between a two-ring molecule vs a single, functionalized ring. For indene and benzonitrile, which both contain a single six-membered ring, it is not necessarily obvious that the addition of a five-membered ring would occur more than the addition of a cyano group, especially given the efficiency of CN addition reactions. A similar argument could be made for additions to a five-membered ring (cyclopentadiene, \ce{C5H6}) to form indene and the cyanocyclopentadienes. This could imply that the conversion from single to multiple-ring molecules may be highly efficient in dark clouds or that indene production does not necessarily directly require the formation of benzene, as is consistent with the proposed pathway by \citet{Doddipatla:2021}. It is also possible that this is related to closed rings' additional vibrational modes to dissipate energy from dissociative processes, making them more stable in interstellar environments.

\begin{table*}[hbt]
    \centering
    \caption{Derived column density measurements for the six ringed molecules detected in the GOTHAM DR2 survey in ascending order of number of heavy atoms.}
    \begin{tabular}{ c c r@{.}l r@{.}l c}
    \toprule
    	Name & Formula &	\multicolumn{2}{c}{$N_T$}     & \multicolumn{2}{c}{$T_{\text{ex}}$} & $\Delta V$		\\
		 & &	\multicolumn{2}{c}{(10$^{11}$\,cm$^{-2}$)}  & \multicolumn{2}{c}{(K)}             & km\,s$^{-1}$\\ 
			\midrule
1-cyanocyclopentadiene & \ce{1-C5H5CN}  & 8&27$^{+0.9}_{-1.0}$ &	6&00$^{+0.03}_{-0.03}$	&	0.119$^{+0.009}_{-0.010}$ \\ 
2-cyanocyclopentadiene & \ce{2-C5H5CN} & 1&89$^{+0.18}_{-0.15}$ & 6&00$^{+0.03}_{-0.03}$	&	0.122$^{+0.010}_{-0.010}$\\
benzonitrile & \ce{C6H5CN} & 17&3$^{+8.5}_{-10}$ & 8&9$^{+0.4}_{-0.4}$	&	0.125$^{+0.005}_{-0.004}$\\
indene & \ce{C9H8} & 96&0$^{+43.3}_{-15.7}$ & 8&55$^{+0.22}_{-0.23}$    & 0.121$^{+0.016}_{-0.016}$\\
1-cyanonaphthalene & \ce{1-C10H7CN} & 7&35$^{+3.30}_{-4.63}$ & 8&9$^{+0.3}_{-0.4}$	&	0.126$^{+0.010}_{-0.009}$\\
2-cyanonaphthalene & \ce{2-C10H7CN} & 7&05$^{+4.50}_{-3.23}$ & 8&7$^{+0.4}_{-0.4}$	&	0.125$^{+0.010}_{-0.009}$\\ 
    \bottomrule
    \end{tabular}
    \label{tab:aromatic cols}
\end{table*}

Regarding the other fitted physical conditions of these six rings, the updated DR2 excitation conditions of benzonitrile, indene, and the cyanonaphthalenes all have fairly consistent excitation temperatures between $T_{\text{ex}}$\,${\sim}$8.5--8.9\,K.  %, while the cyanocyclopentadienes have noticeably lower excitation temperatures by about 2.5--3\,K. 
While the values from the MCMC model can be strongly influenced by the priors (\ce{HC9N} and benzonitrile fits here), the spatial distributions or formation conditions of five-membered rings could differ from those with at least one six-membered ring. All four species were found to be well described by four velocity components as has been previously observed in TMC\nobreakdash-1 \citep{Dobashi:2018,Dobashi:2019}. However, uncertainties in the spatial distribution of these four corresponding velocity components complicates this, providing a strong motivation for higher angular resolution maps of TMC\nobreakdash-1. %to be performed with the VLA, ALMA, LMT, and the GBT.
The line widths are all in agreement within uncertainties, which suggests the overall physical dynamics of the species within TMC\nobreakdash-1 are consistent and quiescent.

\subsection{Implications of Aromatic Chemistry}
\label{sec:implications}

As discussed in Section~\ref{sec:model}, the observed abundance of indene exceed the predictions of astrochemical models by several orders of magnitude, which is consistent with the results found from similar attempts to use primarily gas-phase reactions to produce aromatic molecules in dark clouds. As noted by \citet{Burkhardt:2021}, at least three potential avenues exist that could account for this surprisingly large observed abundance: 1) the existing network is underproducing rings due to missing or underestimated pathways and mechanisms 2) TMC-1 inherited a reservoir of rings and PAHs produced from a top-down chemistry scenario or 3) indene is significantly formed by the active destruction of large PAHs into smaller ones in dark cloud conditions.

For the first avenue, possible promising additions to the network include the cyclization of large carbon-chains, efficient grain-surface processes alongside non-thermal desorption routes, and an incomplete network for crucial carbon-chain precursors. This potential insufficent network may be improved by the recent detections of many partially-saturated carbon-chains which will provide more observational constraints for pathways within astrochemical models \citep{Xue:2020aa,McGuire:2020bb,Loomis:2021aa,Shingledecker:2021}. Indeed, the expansion of the network for smaller semi-saturated carbon chains in this work did somewhat increase the predicted abundances of the other cyclic species (cyclopentadiene by a factor of $\sim$50, benzene by $\sim$0.4, and naphthalene by $\sim$5), although not enough to reproduce the observed gas-phase abundances of their functionalized forms.

A more robust treatment of grain-surface processing could reveal efficient formation pathways for aromatic molecules in cold environments, provided efficient non-thermal desorption processes are included \citep{Shingledecker:2018fm, Burkhardt:2019gf}.  It should be noted that these models did not include the additional solid-phase cosmic ray-driven and suprathermal chemistry used in \citep{Doddipatla:2021} and originally developed in \citet{Shingledecker:2018fm,Shingledecker:2020}. \citet{Shingledecker:2021} provides compelling evidence that the saturation of carbon chains on grain surfaces in dark clouds could be crucial to aromatic production. Therefore, future work is needed to incorporate these processes into chemical networks. In addition, the low-temperature product-branching-ratios for the key reactions leading to these aromatic molecules are currently unknown and are critical to improving existing models. 

Finally, while the peak simulated gas-phase abundances of PAHs are relatively similar (3--6$\times$10$^{-14}$), the proposed pathways differ non-trivially. Here, indene is produced from styrene and toluene, which are themselves formed from semi-saturated carbon chains. Meanwhile, the naphthalenes are produced by reactions of the phenyl radical (\ce{c-C6H5}), a dissociation product of benzene.%, and vinylacetylene (\ce{CH2CHC2H}) or 1,3-butadiene (\ce{CH2CHCHCH2}) \citep{McGuire:2021aa}.
 This difference may explain why the simulated abundance of indene takes twice as long as naphthalene to reach within an order of magnitude of its peak abundance, as the formation of benzene occurs much faster than that of the long semi-saturated carbon chains (e.g. \ce{C6H8}).

The second proposed scenario for the observed overabundance of indene, that these rings are abundant prior to the formation of dark clouds, is similar to that discussed for the previously detected rings \citep{Burkhardt:2021,McCarthy:2020aa,McGuire:2021aa}. Small PAHs ($<$20-30 atoms) should be efficiently destroyed in diffuse clouds by UV photons \citep{chabot_coulomb_220,rapacioli_formation_2006,montillaud_evolution_2013}. However, we can still consider a model starting with an initial indene reservoir. For the previously studied cyclic species, an initial abundance equivalent to a unrealistically large fraction of the total carbon budget (0.5--60\%) was required to reproduce the observed abundances, which is also found to be true for indene. Thus, we find it unlikely that indene was directly inherited from PAH formation that is known to occur in the envelopes of AGB stars. 

However, this does not account for the active destruction of larger PAHs, that are likely present in dark clouds but presumably formed in ABG stars well beforehand, which could be an important source of small aromatics. The efficiency and products of these dissociative processes, be they by chemical reactions or high energy particles, are not well constrained for low-UV photon environments such as TMC-1. This provides a strong motivation to improve our understanding on the survivability of both large and small PAHs to better account for this new detection. Indeed, indene and benzene will have different susceptibilities to photoabsorption, photoexcitation, and photodissociation, which could produce different indene to benzene ratios based on top-down vs bottom-up predictions. As such, while neither the top-down nor the bottom-up scenario can reproduce the observed abundance in these models, we cannot definitely rule out either theory as we have not fully explored the possible pathways each would suggest.

\subsection{Future Molecular Searches}
\label{sec:future}

The growing number of detections of cyclic molecules in GOTHAM and ARKHAM imply that aromatic chemistry is ubiquitous \citep{Burkhardt:2021}, abundant ($N$\,$>$10$^{11}$\,cm$^{-2}$), and revealing an unprecedented degree of interstellar chemical complexity. Indeed, cyclic molecules constitute six of the seven $\geq$12-atom molecules detected in GOTHAM and nine of the ten $\geq$12-atom interstellar molecules detected to date when considering the fullerenes. These three findings all point to the importance of rings and aromatics in the greater context of prestellar chemistry.

Going forward, the potential detection of cyanoindene or purely hydrocarbon counterparts of the detected CN-substituted rings (e.g. cyclopentadiene, $\mu_D\sim$0.42\,D \citep{laurie:635,damiani:265}) will provide key constrains on the bottom-up formation of PAHs, including those that are not detectable with radio astronomy (e.g. benzene and naphthalene). In addition, the importance of the proposed formation pathway of indene would be significantly constrained by the detection of the precursor functionalized aromatics such as styrene and toluene. Similarly, existing pathways to form rings in astrochemical models rely on specific semi-saturated carbon-chain precursors (see \citealt{Burkhardt:2021}, \citealt{Shingledecker:2021}, and Section~\ref{sec:model}), many with little to no observational constraints. Therefore, observational searches should also be focused on filling in these gaps in our knowledge, such as 1,3-butadiene, \ce{C4H8}, \ce{C5H8}, \ce{C6H8}, or their cyano-derivatives when they have small or zero dipole moment.

\section{Conclusions}
\label{sec:con}
\begin{enumerate}
    \item Here, we report a velocity stacked detection of the first pure hydrocarbon PAH, indene toward TMC\nobreakdash-1 as part of the GOTHAM survey.
    \item This detection opens up the possibility to constrain the abundance of aromatic molecules with no permanent dipole by comparing them to their CN-added counterparts, motivating the search for cyclopentadiene and the cyano-indenes.
    \item MCMC modeling derives a column density that implies indene is the most abundant detected molecule containing a five- or six-membered ring in TMC\nobreakdash-1, which may have implications on the efficiency of the formation of multi-ring species.
    \item Similar to the previous cyclic molecules detected in GOTHAM, the observed abundances are several orders of magnitude larger than what chemical models predict, providing motivation to explore additional gas-phase and grain-surface pathways and processes and the production of small rings from the destruction of larger PAHs. 
    \item Further searches for cyano/pure ring counterparts, substituted aromatic rings, and semi-saturated carbon-chain precursors will be vital to constrain the formation of five- and six-membered rings in the interstellar medium.
\end{enumerate}

\section{Data access \& code}

Data used for the MCMC analysis can be found in the DataVerse entry \citep{DVN/K9HRCK_2020}. The code used to perform the analysis is part of the \texttt{molsim} open-source package; an archival version of the code can be accessed at \cite{lee_molsim_2020}.

\facilities{GBT}


\software{
    \texttt{Nautilus} v1.1 \citep{Ruaud:2016},
    \texttt{Molsim} \citep{lee_molsim_2020},
    \texttt{Emcee} \citep{foreman-mackey_emcee_2013},
    \texttt{ArViz} \citep{kumar_arviz_2019}
          }

\emph{Note Added in Proofs}: Some weeks after this manuscript was submitted, a manuscript from Cernicharo et al. was published on the arXiv \citep{Cernicharo:2021b} describing a detection of indene in the same source using the Yebes 40-m telescope at higher frequencies (31.0--50.3\,GHz) but lower spectral resolution (38.15\,kHz).  The derived total column density is in good agreement with our own, considering the differences in beam size.  We also note that the spectroscopic parameters they derived for indene from the Yebes 40-m observations are in line with those we derived and presented here based on our high-resolution laboratory measurements.

\acknowledgments

A.M.B. acknowledges support from the Smithsonian Institution as a Submillimeter Array (SMA) Fellow. A.M.B. and C.N.S. would like to also thank V. Wakelam for use of the \texttt{nautilus} v1.1 code. K.L.K.L., P.B.C., and M.C.M. acknowledge NSF grant AST-1908576 and NASA grant 80NSSC18K0396 for support. I.R.C. acknowledges funding from the European Union’s Horizon 2020 research and innovation programme under the Marie Skłodowska-Curie grant agreement No 845165-MIRAGE. S.B.C. is supported by the NASA Astrobiology Institute through the Goddard Center for Astrobiology. E.H. thanks the National Science Foundation for support through grant AST-1906489. The National Radio Astronomy Observatory is a facility of the National Science Foundation operated under cooperative agreement by Associated Universities, Inc.  The Green Bank Observatory is a facility of the National Science Foundation operated under cooperative agreement by Associated Universities, Inc. The authors would also like to thank the two anonymous reviewers whose comments and feedback substantially improved the manuscript.

%TC:ignore

\bibliography{bibliography}
\bibliographystyle{aasjournal}

\appendix

\renewcommand{\thefigure}{A\arabic{figure}}
\renewcommand{\thetable}{A\arabic{table}}
\renewcommand{\theequation}{A\arabic{equation}}
\setcounter{figure}{0}
\setcounter{table}{0}
\setcounter{equation}{0}

%\begin{landscape}

\begin{figure*}[h]
    \centering
    \includegraphics[width=\textwidth]{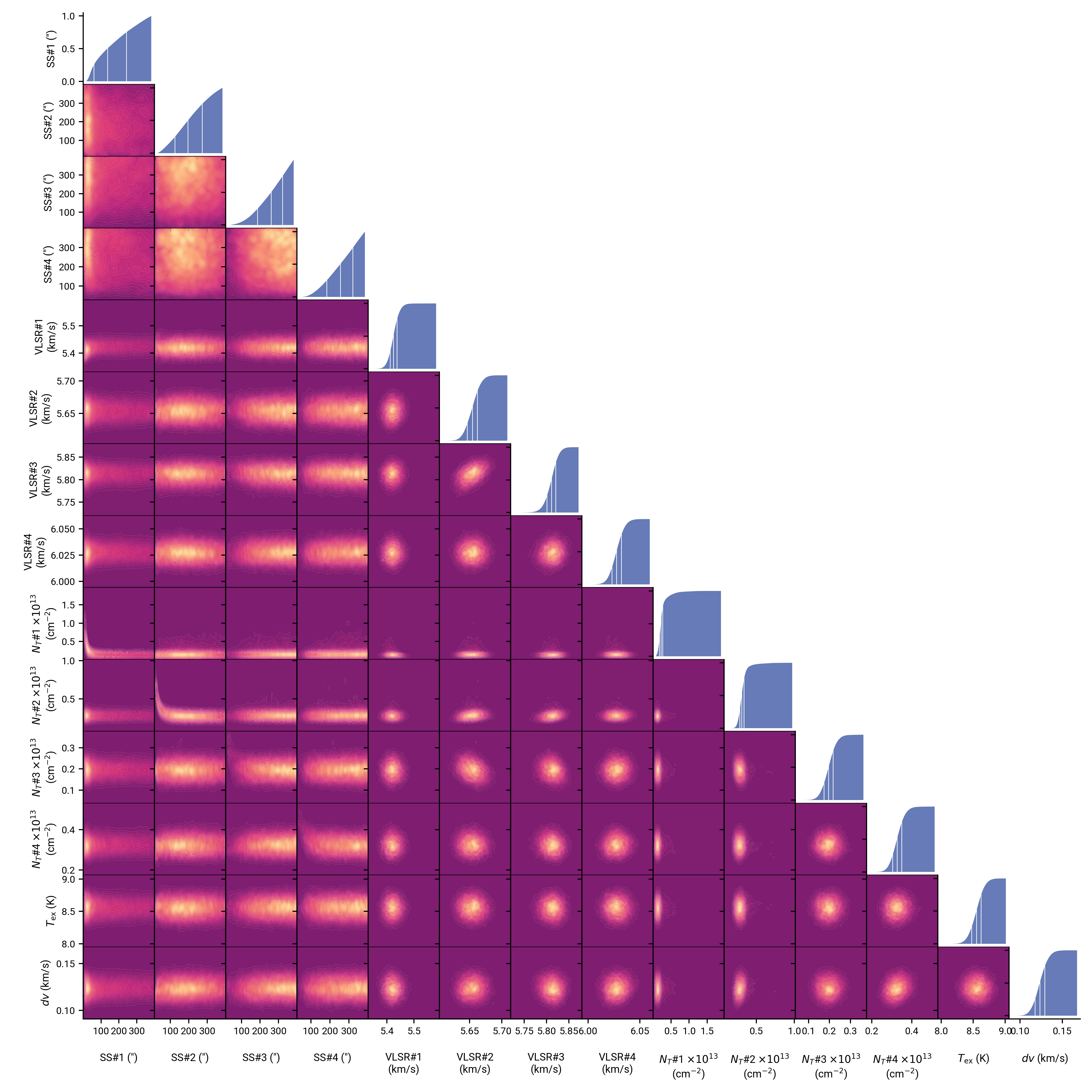}
    \caption{Corner plot for indene. The diagonal traces correspond to ECDF plots, and off-diagonal plots show the kernel density covariance between model parameters. In the former, lines represent the 25$^{\text{th}}$, 50$^{\text{th}}$, and 75$^{\text{th}}$ percentiles respectively. The length scale for the kernel density plots is chosen with Scott's rule.}
    \label{fig:indene_corner}
\end{figure*}

\small
\newcounter{magicrownumbers}
\newcommand\rownumber{\stepcounter{magicrownumbers}\arabic{magicrownumbers}}
\setcounter{magicrownumbers}{0}
\setcounter{magicrownumbers}{-178}

\begin{longrotatetable}
\begin{deluxetable*}{ r l c c c c c c c c c}
    \tablecaption{Indene-related Reactions Added to GOTHAM Network}
    \tablehead{
    \multicolumn2c{Reaction}  & \colhead{$\alpha$} & \colhead{$\beta$} & \colhead{$\gamma$} & \colhead{$T_{\text{min}}$} & \colhead{$T_{\text{max}}$} &  \colhead{Reaction Type} & \colhead{Reference(s)}
    & \colhead{Reaction ID}}
    \startdata
%\ce{C2H} + \ce{C5H8} &$\rightarrow$  \ce{C6H5CH3} + \ce{H} & 3.0$\times10^{-10}$ &  0 & 0 & 3  & \cite{Doddipatla:2021}  & \rownumber \\
%\ce{CCCH3} + \ce{CH2CHCHCH2} &$\rightarrow$ \ce{H} + \ce{C6H5CH3} & 3.0$\times10^{-10}$ &  0 & 0  &3  & \cite{Belisario-Lara:2018}  & \rownumber \\
%\ce{CH} + \ce{C6H5CH3} &$\rightarrow$ \ce{H} + \ce{C6H5C2H3} & 3.0$\times10^{-10}$ &  0 & 0  &3  & \cite{Doddipatla:2021}  & \rownumber \\
%\ce{CH} + \ce{C6H5C2H3} &$\rightarrow$ \ce{H} + \ce{C9H8}    & 4.0$\times10^{-10}$ &  0 & 0  & 3  & \cite{Doddipatla:2021}  & \rownumber \\
\ce{CH} + \ce{C6H5C2H3} & \hspace{-3mm}$\longrightarrow$ \ce{H} + \ce{C9H8} & 4$\times10^{-10}$ & 0 & 0 & - & - & 3 & \tablenotemark{a} & \newtag{\rownumber}{rx:d21prodi}\\
\ce{CH} + \ce{C6H5CH3} & \hspace{-3mm}$\longrightarrow$ \ce{C6H5C2H3} + \ce{H} & 3$\times10^{-10}$ & 0 & 0 & - & - & 3 & \tablenotemark{a} & \rownumber \\
\ce{CCH} + \ce{C5H8} & \hspace{-3mm}$\longrightarrow$ \ce{C6H5CH3} + \ce{H} & 3$\times10^{-10}$ & 0 & 0 & - & - & 3 & \tablenotemark{a} & \rownumber \\
\ce{CCCH3} + \ce{CH2CHCHCH2} & \hspace{-3mm}$\longrightarrow$ \ce{C6H5CH3} + \ce{H} & 3$\times10^{-10}$ & 0 & 0 & - & - & 3 & \tablenotemark{a} & \newtag{\rownumber}{rx:d21prodf}\\
\ce{C9H8} + \ce{C} & \hspace{-3mm}$\longrightarrow$ \ce{C10H7} + \ce{H} & 1$\times10^{-10}$ & 0 & 0 & - & - & 3 & \tablenotemark{a} & \newtag{\rownumber}{rx:d21desti}\\
\ce{C9H8} + \ce{C2} & \hspace{-3mm}$\longrightarrow$ \ce{C11H7} + \ce{H} & 1$\times10^{-10}$ & 0 & 0 & - & - & 3 & \tablenotemark{a} & \rownumber \\
\ce{C9H8} + \ce{CN} & \hspace{-3mm}$\longrightarrow$ \ce{C9H7CN} + \ce{H} & 1$\times10^{-10}$ & 0 & 0 & - & - & 3 & \tablenotemark{a} & \rownumber \\
\ce{C9H8} + \ce{CCH} & \hspace{-3mm}$\longrightarrow$ \ce{C11H8} + \ce{H} & 1$\times10^{-10}$ & 0 & 0 & - & - & 3 & \tablenotemark{a} & \rownumber \\
\ce{C9H8} + \ce{CH} & \hspace{-3mm}$\longrightarrow$ \ce{C10H8} + \ce{H} & 1$\times10^{-10}$ & 0 & 0 & - & - & 3 & \tablenotemark{a} & \rownumber \\
\ce{C6H8} + \ce{CCH} & \hspace{-3mm}$\longrightarrow$ \ce{C6H5C2H3} + \ce{H} & 1$\times10^{-10}$ & 0 & 0 & - & - & 3 & \tablenotemark{a} & \rownumber \\
\ce{C5H8} + \ce{CH} & \hspace{-3mm}$\longrightarrow$ \ce{C6H8} + \ce{H} & 1$\times10^{-10}$ & 0 & 0 & - & - & 3 & \tablenotemark{a} & \newtag{\rownumber}{rx:d21destf}\\
\hline
\ce{C3H7} + \ce{CRP} & \hspace{-3mm}$\longrightarrow$ \ce{H} + \ce{CH3CHCH2} & 5000 & 0 & 0 & - & - & 1 & \tablenotemark{b} & \newtag{\rownumber}{rx:kidai}\\
\ce{C3H7} + \ce{Photon} & \hspace{-3mm}$\longrightarrow$ \ce{H} + \ce{CH3CHCH2} & 1$\times10^{-9}$ & 0 & 1.7 & - & - & 2 & \tablenotemark{b} & \rownumber \\
\ce{C3H8} + \ce{CRP} & \hspace{-3mm}$\longrightarrow$ \ce{H} + \ce{C3H7} & 5000 & 0 & 0 & - & - & 1 & \tablenotemark{b} & \rownumber \\
\ce{C3H8} + \ce{Photon} & \hspace{-3mm}$\longrightarrow$ \ce{H} + \ce{C3H7} & 1$\times10^{-9}$ & 0 & 1.7 & - & - & 2 & \tablenotemark{b} & \rownumber \\
\hline
\ce{H} + \ce{C3H7} & \hspace{-3mm}$\longrightarrow$ \ce{CH3} + \ce{C2H5} & 9.7$\times10^{-11}$ & 0.22 & 0 & 10 & 800 & 3 & \tablenotemark{c},\tablenotemark{d} & \rownumber \\
                   & \hspace{-3mm}$\longrightarrow$ \ce{H2} + \ce{CH3CHCH2} & 3$\times10^{-11}$ & 0 & 0 & 10 & 800 & 3 & \tablenotemark{c},\tablenotemark{d} & \rownumber \\
\ce{C} + \ce{C3H7} & \hspace{-3mm}$\longrightarrow$ \ce{H} + \ce{CH2CHCHCH2} & 1.6$\times10^{-10}$ & 0 & 0 & 10 & 800 & 3 & \tablenotemark{c} & \rownumber \\
                   & \hspace{-3mm}$\longrightarrow$ \ce{CH3} + \ce{CH2CCH2} & 4$\times10^{-11}$ & 0 & 0 & 10 & 800 & 3 & \tablenotemark{c} & \rownumber \\
                   & \hspace{-3mm}$\longrightarrow$ \ce{CH4} + \ce{CH2CCH} & 2$\times10^{-10}$ & 0 & 0 & 10 & 800 & 3 & \tablenotemark{b} & \rownumber \\
\ce{N} + \ce{C3H7} & \hspace{-3mm}$\longrightarrow$ \ce{C2H5} + \ce{H2CN} & 1$\times10^{-10}$ & 0 & 0 & 10 & 800 & 3 & \tablenotemark{c} & \rownumber \\
                   & \hspace{-3mm}$\longrightarrow$ \ce{CH3CHO} + \ce{CH3} & 4$\times10^{-11}$ & 0 & 0 & 10 & 800 & 3 & \tablenotemark{c},\tablenotemark{e},\tablenotemark{f} & \rownumber \\
                   & \hspace{-3mm}$\longrightarrow$ \ce{C2H5} + \ce{H2CO} & 2$\times10^{-11}$ & 0 & 0 & 10 & 800 & 3 & \tablenotemark{c},\tablenotemark{e},\tablenotemark{g} & \rownumber \\
\ce{O} + \ce{C3H7} & \hspace{-3mm}$\longrightarrow$ \ce{CH3COCH3} + \ce{H} & 4$\times10^{-11}$ & 0 & 0 & 10 & 800 & 3 & \tablenotemark{c},\tablenotemark{e},\tablenotemark{g} & \rownumber \\
\ce{C3H9+} + \ce{e-} & \hspace{-3mm}$\longrightarrow$ \ce{H} + \ce{C3H8} & 2$\times10^{-7}$ & -0.7 & 0 & 10 & 300 & 3 & \tablenotemark{c},\tablenotemark{h},\tablenotemark{i},\tablenotemark{j},\tablenotemark{k},\tablenotemark{l} & \rownumber \\
                     & \hspace{-3mm}$\longrightarrow$ \ce{C3H7} + \ce{H} + \ce{H} & 2$\times10^{-7}$ & -0.7 & 0 & 10 & 800 & 3 & \tablenotemark{c},\tablenotemark{h},\tablenotemark{i},\tablenotemark{j},\tablenotemark{k},\tablenotemark{l} & \rownumber \\
                     & \hspace{-3mm}$\longrightarrow$ \ce{H} + \ce{H2} + \ce{CH3CHCH2} & 2$\times10^{-7}$ & -0.7 & 0 & 10 & 800 & 3 & \tablenotemark{c},\tablenotemark{h},\tablenotemark{i},\tablenotemark{j},\tablenotemark{l} & \rownumber \\
                     & \hspace{-3mm}$\longrightarrow$ \ce{CH3} + \ce{C2H6} & 1$\times10^{-7}$ & -0.7 & 0 & 10 & 800 & 3 & \tablenotemark{c},\tablenotemark{h},\tablenotemark{i},\tablenotemark{j},\tablenotemark{l} & \rownumber \\
                     & \hspace{-3mm}$\longrightarrow$ \ce{H2} + \ce{CH3} + \ce{C2H4} & 1$\times10^{-7}$ & -0.7 & 0 & 10 & 800 & 3 & \tablenotemark{c},\tablenotemark{h},\tablenotemark{i},\tablenotemark{j},\tablenotemark{l} & \rownumber \\
\ce{CH2} + \ce{C3H7} & \hspace{-3mm}$\longrightarrow$ \ce{C2H4} + \ce{C2H5} & 3$\times10^{-11}$ & 0 & 0 & 50 & 200 & 3 & \tablenotemark{g},\tablenotemark{m} & \rownumber \\
                     & \hspace{-3mm}$\longrightarrow$ \ce{CH3} + \ce{CH3CHCH2} & 3$\times10^{-12}$ & 0 & 0 & 50 & 200 & 3 & \tablenotemark{g},\tablenotemark{m} & \rownumber \\
\ce{CH3} + \ce{C3H7} & \hspace{-3mm}$\longrightarrow$ \ce{CH3CHCH2} + \ce{CH4} & 3.1$\times10^{-12}$ & -0.32 & 0 & 50 & 200 & 3 & \tablenotemark{g},\tablenotemark{m} & \rownumber \\
\ce{CCH} + \ce{C3H7} & \hspace{-3mm}$\longrightarrow$ \ce{C2H2} + \ce{CH3CHCH2} & 1$\times10^{-11}$ & 0 & 0 & 50 & 200 & 3 & \tablenotemark{g},\tablenotemark{m} & \rownumber \\
                     & \hspace{-3mm}$\longrightarrow$ \ce{CH2CCH} + \ce{C2H5} & 2$\times10^{-11}$ & 0 & 0 & 50 & 200 & 3 & \tablenotemark{g},\tablenotemark{m} & \rownumber \\
\ce{C2H3} + \ce{C3H7} & \hspace{-3mm}$\longrightarrow$ \ce{CH3CHCH2} + \ce{C2H4} & 2$\times10^{-12}$ & 0 & 0 & 50 & 200 & 3 & \tablenotemark{g},\tablenotemark{m} & \rownumber \\
                      & \hspace{-3mm}$\longrightarrow$ \ce{C2H2} + \ce{C3H8} & 2$\times10^{-12}$ & 0 & 0 & 50 & 200 & 3 & \tablenotemark{g},\tablenotemark{m} & \rownumber \\
\ce{C2H5} + \ce{C3H7} & \hspace{-3mm}$\longrightarrow$ \ce{C2H4} + \ce{C3H8} & 1.9$\times10^{-12}$ & 0 & 0 & 50 & 200 & 3 & \tablenotemark{g},\tablenotemark{m} & \rownumber \\
                      & \hspace{-3mm}$\longrightarrow$ \ce{CH3CHCH2} + \ce{C2H6} & 2.4$\times10^{-12}$ & 0 & 0 & 50 & 200 & 3 & \tablenotemark{g},\tablenotemark{m} & \rownumber \\
\ce{C3H5} + \ce{C3H7} & \hspace{-3mm}$\longrightarrow$ \ce{CH3CHCH2} + \ce{CH3CHCH2} & 2.4$\times10^{-12}$ & 0 & -66 & 50 & 200 & 3 & \tablenotemark{g},\tablenotemark{m} & \rownumber \\
                      & \hspace{-3mm}$\longrightarrow$ \ce{C3H8} + \ce{CH2CCH2} & 1.2$\times10^{-12}$ & 0 & -66 & 50 & 200 & 3 & \tablenotemark{g},\tablenotemark{m} & \rownumber \\
\ce{H2} + \ce{C3H7} & \hspace{-3mm}$\longrightarrow$ \ce{H} + \ce{C3H8} & 3.3$\times10^{-14}$ & 2.8 & 4600 & 50 & 200 & 3 & \tablenotemark{g},\tablenotemark{m} & \rownumber \\
\ce{CH4} + \ce{C3H7} & \hspace{-3mm}$\longrightarrow$ \ce{CH3} + \ce{C3H8} & 3.6$\times10^{-16}$ & 4 & 5470 & 50 & 200 & 3 & \tablenotemark{g},\tablenotemark{m} & \rownumber \\
\ce{C2H2} + \ce{C3H7} & \hspace{-3mm}$\longrightarrow$ \ce{C2H4} + \ce{C3H5} & 1.2$\times10^{-12}$ & 0 & 4530 & 50 & 200 & 3 & \tablenotemark{g},\tablenotemark{m} & \rownumber \\
\ce{C2H6} + \ce{C3H7} & \hspace{-3mm}$\longrightarrow$ \ce{C2H5} + \ce{C3H8} & 1.2$\times10^{-15}$ & 3.8 & 4550 & 50 & 200 & 3 & \tablenotemark{g},\tablenotemark{m} & \rownumber \\
\ce{C3H7} + \ce{C3H7} & \hspace{-3mm}$\longrightarrow$ \ce{CH3CHCH2} + \ce{C3H8} & 2.8$\times10^{-12}$ & 0 & 0 & 50 & 200 & 3 & \tablenotemark{g},\tablenotemark{m} & \rownumber \\
\ce{HCO} + \ce{C3H7} & \hspace{-3mm}$\longrightarrow$ \ce{CO} + \ce{C3H8} & 1$\times10^{-10}$ & 0 & 0 & 50 & 200 & 3 & \tablenotemark{g},\tablenotemark{m} & \rownumber \\
\ce{C2H4} + \ce{C3H7} & \hspace{-3mm}$\longrightarrow$ \ce{C2H3} + \ce{C3H8} & 5.7$\times10^{-14}$ & 3.1 & 9060 & 50 & 200 & 3 & \tablenotemark{g} & \rownumber \\
\ce{C4H} + \ce{C3H7} & \hspace{-3mm}$\longrightarrow$ \ce{C2H5} + \ce{C5H3} & 2$\times10^{-11}$ & 0 & 0 & 50 & 200 & 3 & \tablenotemark{g} & \rownumber \\
                     & \hspace{-3mm}$\longrightarrow$ \ce{CH3CHCH2} + \ce{C4H2} & 1$\times10^{-11}$ & 0 & 0 & 50 & 200 & 3 & \tablenotemark{g} & \rownumber \\
\ce{C4H3} + \ce{C3H7} & \hspace{-3mm}$\longrightarrow$ \ce{CH3CHCH2} + \ce{CH2CHC2H} & 2$\times10^{-12}$ & 0 & 0 & 50 & 200 & 3 & \tablenotemark{g} & \rownumber \\
                      & \hspace{-3mm}$\longrightarrow$ \ce{C4H2} + \ce{C3H8} & 2$\times10^{-12}$ & 0 & 0 & 50 & 200 & 3 & \tablenotemark{g} & \rownumber \\
\ce{CH3CHCH2} + \ce{C3H7} & \hspace{-3mm}$\longrightarrow$ \ce{C3H5} + \ce{C3H8} & 1.7$\times10^{-15}$ & 3.5 & 3340 & 50 & 200 & 3 & \tablenotemark{g},\tablenotemark{n} & \rownumber \\
\hline
\ce{C+} + \ce{C3H8} & \hspace{-3mm}$\longrightarrow$ \ce{C2H5+} + \ce{C2H3} & 7$\times10^{-10}$ & 0 & 0 & 10 & 800 & 3 & \tablenotemark{c},\tablenotemark{o} & \rownumber \\
                    & \hspace{-3mm}$\longrightarrow$ \ce{C2H3+} + \ce{C2H5} & 5$\times10^{-10}$ & 0 & 0 & 10 & 800 & 3 & \tablenotemark{c},\tablenotemark{o} & \rownumber \\
                    & \hspace{-3mm}$\longrightarrow$ \ce{C2H2+} + \ce{C2H6} & 4$\times10^{-10}$ & 0 & 0 & 10 & 800 & 3 & \tablenotemark{c},\tablenotemark{o} & \rownumber \\
\ce{H3+} +  \ce{C3H8} & \hspace{-3mm}$\longrightarrow$ \ce{H2} + \ce{C3H9+} & 1 & 3.3$\times10^{-9}$ & 0 & 10 & 800 & 5 & \tablenotemark{c} & \rownumber \\
\ce{HCO+} + \ce{C3H8} & \hspace{-3mm}$\longrightarrow$ \ce{CO} + \ce{C3H9+} & 1 & 1.3$\times10^{-9}$ & 0 & 10 & 800 & 5 & \tablenotemark{c} & \rownumber \\
\ce{H3O+} + \ce{C3H8} & \hspace{-3mm}$\longrightarrow$ \ce{H2O} + \ce{C3H9+} & 1 & 1.5$\times10^{-9}$ & 0 & 10 & 800 & 5 & \tablenotemark{c} & \rownumber \\
\ce{CH3+} + \ce{C3H8} & \hspace{-3mm}$\longrightarrow$ \ce{CH4} + \ce{C3H7+} & 1$\times10^{-9}$ & 0 & 0 & 10 & 300 & 3 & \tablenotemark{c},\tablenotemark{o} & \rownumber \\
\ce{C2H5+} + \ce{C3H8} & \hspace{-3mm}$\longrightarrow$ \ce{C2H6} + \ce{C3H7+} & 6.3$\times10^{-10}$ & 0 & 0 & 300 & 300 & 3 & \tablenotemark{p},\tablenotemark{q} & \rownumber \\
\ce{N2+} + \ce{C3H8} & \hspace{-3mm}$\longrightarrow$ \ce{H} + \ce{H2} + \ce{N2} + \ce{C3H5+} & 1.7$\times10^{-10}$ & 0 & 0 & 298 & 298 & 3 & \tablenotemark{p} & \rownumber \\
                     & \hspace{-3mm}$\longrightarrow$ \ce{N2} + \ce{CH4} + \ce{C2H4+} & 2.2$\times10^{-10}$ & 0 & 0 & 298 & 298 & 3 & \tablenotemark{p} & \rownumber \\
                     & \hspace{-3mm}$\longrightarrow$ \ce{H} + \ce{N2} + \ce{CH4} + \ce{C2H3+} & 5.2$\times10^{-10}$ & 0 & 0 & 298 & 298 & 3 & \tablenotemark{p} & \rownumber \\
                     & \hspace{-3mm}$\longrightarrow$ \ce{N2} + \ce{CH3} + \ce{C2H5+} & 3.9$\times10^{-10}$ & 0 & 0 & 298 & 298 & 3 & \tablenotemark{p} & \rownumber \\
\ce{H} + \ce{C3H8} & \hspace{-3mm}$\longrightarrow$ \ce{H2} + \ce{C3H7} & 4.3$\times10^{-12}$ & 2.5 & 3400 & 50 & 200 & 3 & \tablenotemark{g},\tablenotemark{m} & \rownumber \\
\ce{CH2} + \ce{C3H8} & \hspace{-3mm}$\longrightarrow$ \ce{CH3} + \ce{C3H7} & 1.6$\times10^{-15}$ & 3.6 & 3600 & 50 & 200 & 3 & \tablenotemark{g},\tablenotemark{m} & \rownumber \\
\ce{CH3} + \ce{C3H8} & \hspace{-3mm}$\longrightarrow$ \ce{CH4} + \ce{C3H7} & 1.6$\times10^{-15}$ & 3.6 & 3600 & 50 & 200 & 3 & \tablenotemark{g},\tablenotemark{m} & \rownumber \\
\ce{CCH} + \ce{C3H8} & \hspace{-3mm}$\longrightarrow$ \ce{C2H2} + \ce{C3H7} & 9.8$\times10^{-11}$ & 0 & 71 & 50 & 200 & 3 & \tablenotemark{g},\tablenotemark{m} & \rownumber \\
\ce{C2H3} + \ce{C3H8} & \hspace{-3mm}$\longrightarrow$ \ce{C2H4} + \ce{C3H7} & 5$\times10^{-16}$ & 2.3 & 5280 & 50 & 200 & 3 & \tablenotemark{g},\tablenotemark{m} & \rownumber \\
\ce{C2H5} + \ce{C3H8} & \hspace{-3mm}$\longrightarrow$ \ce{C2H6} + \ce{C3H7} & 1.6$\times10^{-15}$ & 3.6 & 4600 & 50 & 200 & 3 & \tablenotemark{g},\tablenotemark{m} & \rownumber \\
\ce{HCO} + \ce{C3H8} & \hspace{-3mm}$\longrightarrow$ \ce{H2CO} + \ce{C3H7} & 5.3$\times10^{-13}$ & 2.5 & 9290 & 50 & 200 & 3 & \tablenotemark{g},\tablenotemark{m} & \rownumber \\
\ce{CH2CCH} + \ce{C3H8} & \hspace{-3mm}$\longrightarrow$ \ce{CH3CCH} + \ce{C3H7} & 5.8$\times10^{-14}$ & 3.3 & 9990 & 50 & 200 & 3 & \tablenotemark{g} & \rownumber \\
\ce{CH2CCH} + \ce{C3H8} & \hspace{-3mm}$\longrightarrow$ \ce{C3H7} + \ce{CH2CCH2} & 5.8$\times10^{-14}$ & 3.3 & 9990 & 50 & 200 & 3 & \tablenotemark{g} & \rownumber \\
\ce{C3H5} + \ce{C3H8} & \hspace{-3mm}$\longrightarrow$ \ce{CH3CHCH2} + \ce{C3H7} & 5.8$\times10^{-14}$ & 3.3 & 9990 & 50 & 200 & 3 & \tablenotemark{g},\tablenotemark{n} & \rownumber \\
\ce{CH} + \ce{C3H8} & \hspace{-3mm}$\longrightarrow$ \ce{H} + \ce{C4H8} & 1.9$\times10^{-10}$ & 0 & -240 & 50 & 200 & 3 & \tablenotemark{g},\tablenotemark{r} & \rownumber \\
\ce{C4H} + \ce{C3H8} & \hspace{-3mm}$\longrightarrow$ \ce{C4H2} + \ce{C3H7} & 1$\times10^{-10}$ & -1.4 & 56 & 50 & 200 & 3 & \tablenotemark{g},\tablenotemark{s} & \rownumber \\
\ce{C4H3} + \ce{C3H8} & \hspace{-3mm}$\longrightarrow$ \ce{CH2CHC2H} + \ce{C3H7} & 5$\times10^{-16}$ & 2.3 & 5280 & 50 & 200 & 3 & \tablenotemark{g} & \rownumber \\
\ce{CN} + \ce{C3H8} & \hspace{-3mm}$\longrightarrow$ \ce{HCN} + \ce{C3H7} & 8$\times10^{-11}$ & -0.4 & 0 & 10 & 300 & 3 & \tablenotemark{t} & \rownumber \\
\hline
\ce{C2H4} + \ce{C3H5} & \hspace{-3mm}$\longrightarrow$ \ce{H} + \ce{C5H8} & 1$\times10^{-14}$ & 0 & 5780 & 50 & 200 & 3 & \tablenotemark{g},\tablenotemark{u} & \rownumber \\
\ce{CH} + \ce{C4H8} & \hspace{-3mm}$\longrightarrow$ \ce{H} + \ce{C5H8} & 4.3$\times10^{-10}$ & -0.53 & 34 & 50 & 200 & 3 & \tablenotemark{g},\tablenotemark{u} & \rownumber \\
\ce{CCH}+ \ce{C4H8} & \hspace{-3mm}$\longrightarrow$ \ce{H} + \ce{C6H8} & 2.1$\times10^{-10}$ & 0 & 0 & 50 & 200 & 3 & \tablenotemark{g},\tablenotemark{v} & \rownumber \\
\ce{C2H3} + \ce{CH3CHCH2} & \hspace{-3mm}$\longrightarrow$ \ce{C2H4} + \ce{C3H5} & 1.7$\times10^{-15}$ & 3.5 & 2360 & 50 & 200 & 3 & \tablenotemark{g},\tablenotemark{n} & \rownumber \\
                          & \hspace{-3mm}$\longrightarrow$ \ce{H} + \ce{C5H8} & 1.2$\times10^{-12}$ & 0 & 3240 & 50 & 200 & 3 & \tablenotemark{g},\tablenotemark{n} & \newtag{\rownumber}{rx:kidaf}\\
\hline
%\ce{C9H8} + \ce{H3+} & \hspace{-3mm}$\longrightarrow$ \ce{C6H5+} + \ce{CH2CCH2} + \ce{H2} & 0.33 & 5.1$\times10^{-9}$ & 0.66 & 10 & 800 & 4 & \tablenotemark{w} & \rownumber \\
%                     & \hspace{-3mm}$\longrightarrow$ \ce{C6H5+} + \ce{CH3CCH} + \ce{H2} & 0.33 & 5.1$\times10^{-9}$ & 0.66 & 10 & 800 & 4 & \tablenotemark{w} & \rownumber \\
\ce{C9H8} + \ce{He+} & \hspace{-3mm}$\longrightarrow$ \ce{C6H5+} + \ce{CH2CCH} + \ce{He} & 0.33 & 4.5$\times10^{-9}$ & 0.66 & 10 & 800 & 4 & \tablenotemark{w} & \newtag{\rownumber}{rx:thisi}\\
                     & \hspace{-3mm}$\longrightarrow$ \ce{C6H5+} + \ce{CCCH3} + \ce{He} & 0.33 & 4.5$\times10^{-9}$ & 0.66 & 10 & 800 & 4 & \tablenotemark{w} & \rownumber \\
                     & \hspace{-3mm}$\longrightarrow$ \ce{C5H5+} + \ce{C4H3} + \ce{He} & 0.33 & 4.5$\times10^{-9}$ & 0.66 & 10 & 800 & 4 & \tablenotemark{w} & \rownumber \\
%\ce{C9H8} + \ce{H+} & \hspace{-3mm}$\longrightarrow$ \ce{C6H5+} + \ce{C2H3} + \ce{CH} & 0.17 & 8.9$\times10^{-9}$ & 0.66 & 10 & 800 & 4 & \tablenotemark{w} & \rownumber \\
%                    & \hspace{-3mm}$\longrightarrow$ \ce{C6H5+} + \ce{CH3} + \ce{CCH} & 0.17 & 8.9$\times10^{-9}$ & 0.66 & 10 & 800 & 4 & \tablenotemark{w} & \rownumber \\
%                    & \hspace{-3mm}$\longrightarrow$ \ce{C6H5+} + \ce{C2H2} + \ce{CH2} & 0.17 & 8.9$\times10^{-9}$ & 0.66 & 10 & 800 & 4 & \tablenotemark{w} & \rownumber \\
\ce{C9H8} + \ce{C+} & \hspace{-3mm}$\longrightarrow$ \ce{C6H5+} + \ce{CH2CCH} + \ce{C} & 0.25 & 2.6$\times10^{-9}$ & 0.66 & 10 & 800 & 4 & \tablenotemark{w} & \rownumber \\
                    & \hspace{-3mm}$\longrightarrow$ \ce{C6H5+} + \ce{CCCH3} + \ce{C} & 0.25 & 2.6$\times10^{-9}$ & 0.66 & 10 & 800 & 4 & \tablenotemark{w} & \rownumber \\
                    & \hspace{-3mm}$\longrightarrow$ \ce{C6H5+} + \ce{C4H2} + \ce{H} & 0.25 & 2.6$\times10^{-9}$ & 0.66 & 10 & 800 & 4 & \tablenotemark{w} & \rownumber \\
                    & \hspace{-3mm}$\longrightarrow$ \ce{C5H5+} + \ce{C4H3} + \ce{C} & 0.25 & 2.6$\times10^{-9}$ & 0.66 & 10 & 800 & 4 & \tablenotemark{w} & \rownumber \\
%\ce{C9H8} + \ce{HCO+} & \hspace{-3mm}$\longrightarrow$ \ce{C6H5+} + \ce{CH2CCH2} + \ce{CO} & 0.33 & 1.7$\times10^{-9}$ & 0.66 & 10 & 800 & 4 & \tablenotemark{w} & \rownumber \\
%                      & \hspace{-3mm}$\longrightarrow$ \ce{C6H5+} + \ce{CH3CCH} + \ce{CO} & 0.33 & 1.7$\times10^{-9}$ & 0.66 & 10 & 800 & 4 & \tablenotemark{w} & \rownumber \\
%\ce{C9H8} + \ce{H3+} & \hspace{-3mm}$\longrightarrow$ \ce{C5H5+} + \ce{CH2CHC2H} + \ce{H2} & 0.33 & 5.1$\times10^{-9}$ & 0.66 & 10 & 800 & 4 & \tablenotemark{w} & \rownumber \\
%\ce{C9H8} + \ce{H+} & \hspace{-3mm}$\longrightarrow$ \ce{C5H5+} + \ce{C2H3} + \ce{CCH} & 0.17 & 8.9$\times10^{-9}$ & 0.66 & 10 & 800 & 4 & \tablenotemark{w} & \rownumber \\
%                    & \hspace{-3mm}$\longrightarrow$ \ce{C5H5+} + \ce{CH2CCH} + \ce{CH} & 0.17 & 8.9$\times10^{-9}$ & 0.66 & 10 & 800 & 4 & \tablenotemark{w} & \rownumber \\
%                    & \hspace{-3mm}$\longrightarrow$ \ce{C5H5+} + \ce{CCCH3} + \ce{CH} & 0.17 & 8.9$\times10^{-9}$ & 0.66 & 10 & 800 & 4 & \tablenotemark{w} & \rownumber \\
%\ce{C9H8} + \ce{HCO+} & \hspace{-3mm}$\longrightarrow$ \ce{C5H5+} + \ce{CH2CHC2H} + \ce{CO} & 0.33 & 1.7$\times10^{-9}$ & 0.66 & 10 & 800 & 4 & \tablenotemark{w} & \rownumber \\
\ce{C9H7CN} + \ce{H3+} & \hspace{-3mm}$\longrightarrow$ \ce{C6H5+} + \ce{l-C3H2} + \ce{H2} + \ce{HCN} & 0.2 & 5.6$\times10^{-9}$ & 3.9 & 10 & 800 & 4 & \tablenotemark{w} & \rownumber \\
                       & \hspace{-3mm}$\longrightarrow$ \ce{C6H5+} + \ce{CH2CCH} + \ce{H2} + \ce{CN} & 0.2 & 5.6$\times10^{-9}$ & 3.9 & 10 & 800 & 4 & \tablenotemark{w} & \rownumber \\
                       & \hspace{-3mm}$\longrightarrow$ \ce{C6H5+} + \ce{CCCH3} + \ce{H2} + \ce{CN} & 0.2 & 5.6$\times10^{-9}$ & 3.9 & 10 & 800 & 4 & \tablenotemark{w} & \rownumber \\
                       & \hspace{-3mm}$\longrightarrow$ \ce{C5H5+} + \ce{C4H2} + \ce{H2} + \ce{HCN} & 0.2 & 5.6$\times10^{-9}$ & 3.9 & 10 & 800 & 4 & \tablenotemark{w} & \rownumber \\
                       & \hspace{-3mm}$\longrightarrow$ \ce{C5H5+} + \ce{C4H3} + \ce{H2} + \ce{CN} & 0.2 & 5.6$\times10^{-9}$ & 3.9 & 10 & 800 & 4 & \tablenotemark{w} & \rownumber \\
\ce{C9H7CN} + \ce{He+} & \hspace{-3mm}$\longrightarrow$ \ce{C6H5+} + \ce{l-C3H2} + \ce{He} + \ce{CN} & 0.5 & 4.8$\times10^{-9}$ & 3.9 & 10 & 800 & 4 & \tablenotemark{w} & \rownumber \\
                       & \hspace{-3mm}$\longrightarrow$ \ce{C5H5+} + \ce{C4H2} + \ce{He} + \ce{CN} & 0.5 & 4.8$\times10^{-9}$ & 3.9 & 10 & 800 & 4 & \tablenotemark{w} & \rownumber \\
\ce{C9H7CN} + \ce{H+} & \hspace{-3mm}$\longrightarrow$ \ce{C6H5+} + \ce{l-C3H2} + \ce{HCN} & 0.2 & 9.6$\times10^{-9}$ & 3.9 & 10 & 800 & 4 & \tablenotemark{w} & \rownumber \\
                      & \hspace{-3mm}$\longrightarrow$ \ce{C6H5+} + \ce{CH2CCH} + \ce{CN} & 0.2 & 9.6$\times10^{-9}$ & 3.9 & 10 & 800 & 4 & \tablenotemark{w} & \rownumber \\
                      & \hspace{-3mm}$\longrightarrow$ \ce{C6H5+} + \ce{CCCH3} + \ce{CN} & 0.2 & 9.6$\times10^{-9}$ & 3.9 & 10 & 800 & 4 & \tablenotemark{w} & \rownumber \\
                      & \hspace{-3mm}$\longrightarrow$ \ce{C5H5+} + \ce{C4H2} + \ce{HCN} & 0.2 & 9.6$\times10^{-9}$ & 3.9 & 10 & 800 & 4 & \tablenotemark{w} & \rownumber \\
                      & \hspace{-3mm}$\longrightarrow$ \ce{C5H5+} + \ce{C4H3} + \ce{CN} & 0.2 & 9.6$\times10^{-9}$ & 3.9 & 10 & 800 & 4 & \tablenotemark{w} & \rownumber \\
\ce{C9H7CN} + \ce{C+} & \hspace{-3mm}$\longrightarrow$ \ce{C6H5+} + \ce{l-C3H2} + \ce{CCN} & 0.25 & 2.8$\times10^{-9}$ & 3.9 & 10 & 800 & 4 & \tablenotemark{w} & \rownumber \\
                      & \hspace{-3mm}$\longrightarrow$ \ce{C6H5+} + \ce{C4H2} + \ce{CN} & 0.25 & 2.8$\times10^{-9}$ & 3.9 & 10 & 800 & 4 & \tablenotemark{w} & \rownumber \\
                      & \hspace{-3mm}$\longrightarrow$ \ce{C5H5+} + \ce{C4H2} + \ce{CCN} & 0.25 & 2.8$\times10^{-9}$ & 3.9 & 10 & 800 & 4 & \tablenotemark{w} & \rownumber \\
                      & \hspace{-3mm}$\longrightarrow$ \ce{C5H5+} + \ce{C5H2} + \ce{CN} & 0.25 & 2.8$\times10^{-9}$ & 3.9 & 10 & 800 & 4 & \tablenotemark{w} & \rownumber \\
\ce{C9H7CN} + \ce{HCO+} & \hspace{-3mm}$\longrightarrow$ \ce{C6H5+} + \ce{l-C3H2} + \ce{CO} + \ce{HCN} & 0.2 & 1.8$\times10^{-9}$ & 3.9 & 10 & 800 & 4 & \tablenotemark{w} & \rownumber \\
                        & \hspace{-3mm}$\longrightarrow$ \ce{C6H5+} + \ce{CH2CCH} + \ce{CO} + \ce{CN} & 0.2 & 1.8$\times10^{-9}$ & 3.9 & 10 & 800 & 4 & \tablenotemark{w} & \rownumber \\
                        & \hspace{-3mm}$\longrightarrow$ \ce{C6H5+} + \ce{CCCH3} + \ce{CO} + \ce{CN} & 0.2 & 1.8$\times10^{-9}$ & 3.9 & 10 & 800 & 4 & \tablenotemark{w} & \rownumber \\
                        & \hspace{-3mm}$\longrightarrow$ \ce{C5H5+} + \ce{C4H2} + \ce{CO} + \ce{HCN} & 0.2 & 1.8$\times10^{-9}$ & 3.9 & 10 & 800 & 4 & \tablenotemark{w} & \rownumber \\
                        & \hspace{-3mm}$\longrightarrow$ \ce{C5H5+} + \ce{C4H3} + \ce{CO} + \ce{CN} & 0.2 & 1.8$\times10^{-9}$ & 3.9 & 10 & 800 & 4 & \tablenotemark{w} & \rownumber \\
\hline
\ce{C4H8} + \ce{H3+} & \hspace{-3mm}$\longrightarrow$ \ce{CH4} + \ce{C3H5+} + \ce{H2} & 0.17 & 3.6$\times10^{-9}$ & 0.56 & 10 & 800 & 4 & \tablenotemark{w} & \rownumber \\
                     & \hspace{-3mm}$\longrightarrow$ \ce{CH3+} + \ce{CH3CHCH2} + \ce{H2} & 0.17 & 3.6$\times10^{-9}$ & 0.56 & 10 & 800 & 4 & \tablenotemark{w} & \rownumber \\
                     & \hspace{-3mm}$\longrightarrow$ \ce{C2H6} + \ce{C2H3+} + \ce{H2} & 0.17 & 3.6$\times10^{-9}$ & 0.56 & 10 & 800 & 4 & \tablenotemark{w} & \rownumber \\
                     & \hspace{-3mm}$\longrightarrow$ \ce{C2H5+} + \ce{C2H4} + \ce{H2} & 0.17 & 3.6$\times10^{-9}$ & 0.56 & 10 & 800 & 4 & \tablenotemark{w} & \rownumber \\
                     & \hspace{-3mm}$\longrightarrow$ \ce{CH2+} + \ce{C3H7} + \ce{H2} & 0.17 & 3.6$\times10^{-9}$ & 0.56 & 10 & 800 & 4 & \tablenotemark{w} & \rownumber \\
                     & \hspace{-3mm}$\longrightarrow$ \ce{C4H7+} + \ce{H2} + \ce{H2} & 0.17 & 3.6$\times10^{-9}$ & 0.56 & 10 & 800 & 4 & \tablenotemark{w} & \rownumber \\
\ce{C4H8} + \ce{HCO+} & \hspace{-3mm}$\longrightarrow$ \ce{CH4} + \ce{C3H5+} + \ce{CO} & 0.17 & 1.2$\times10^{-9}$ & 0.56 & 10 & 800 & 4 & \tablenotemark{w} & \rownumber \\
                      & \hspace{-3mm}$\longrightarrow$ \ce{CH3+} + \ce{CH3CHCH2} + \ce{CO} & 0.17 & 1.2$\times10^{-9}$ & 0.56 & 10 & 800 & 4 & \tablenotemark{w} & \rownumber \\
                      & \hspace{-3mm}$\longrightarrow$ \ce{C2H6} + \ce{C2H3+} + \ce{CO} & 0.17 & 1.2$\times10^{-9}$ & 0.56 & 10 & 800 & 4 & \tablenotemark{w} & \rownumber \\
                      & \hspace{-3mm}$\longrightarrow$ \ce{C2H5+} + \ce{C2H4} + \ce{CO} & 0.17 & 1.2$\times10^{-9}$ & 0.56 & 10 & 800 & 4 & \tablenotemark{w} & \rownumber \\
                      & \hspace{-3mm}$\longrightarrow$ \ce{CH2+} + \ce{C3H7} + \ce{CO} & 0.17 & 1.2$\times10^{-9}$ & 0.56 & 10 & 800 & 4 & \tablenotemark{w} & \rownumber \\
                      & \hspace{-3mm}$\longrightarrow$ \ce{C4H7+} + \ce{H2} + \ce{CO} & 0.17 & 1.2$\times10^{-9}$ & 0.56 & 10 & 800 & 4 & \tablenotemark{w} & \rownumber \\
\ce{C4H8} + \ce{C+} & \hspace{-3mm}$\longrightarrow$ \ce{CH3} + \ce{C3H5+} + \ce{C} & 0.2 & 1.8$\times10^{-9}$ & 0.56 & 10 & 800 & 4 & \tablenotemark{w} & \rownumber \\
                    & \hspace{-3mm}$\longrightarrow$ \ce{CH3+} + \ce{C3H5} + \ce{C} & 0.2 & 1.8$\times10^{-9}$ & 0.56 & 10 & 800 & 4 & \tablenotemark{w} & \rownumber \\
                    & \hspace{-3mm}$\longrightarrow$ \ce{C2H5+} + \ce{C2H3} + \ce{C} & 0.2 & 1.8$\times10^{-9}$ & 0.56 & 10 & 800 & 4 & \tablenotemark{w} & \rownumber \\
                    & \hspace{-3mm}$\longrightarrow$ \ce{C2H5} + \ce{C2H3+} + \ce{C} & 0.2 & 1.8$\times10^{-9}$ & 0.56 & 10 & 800 & 4 & \tablenotemark{w} & \rownumber \\
                    & \hspace{-3mm}$\longrightarrow$ \ce{CH2+} + \ce{CH3CHCH2} + \ce{C} & 0.2 & 1.8$\times10^{-9}$ & 0.56 & 10 & 800 & 4 & \tablenotemark{w} & \rownumber \\
\ce{C4H8} + \ce{He+} & \hspace{-3mm}$\longrightarrow$ \ce{CH3} + \ce{C3H5+} + \ce{He} & 0.2 & 3.1$\times10^{-9}$ & 0.56 & 10 & 800 & 4 & \tablenotemark{w} & \rownumber \\
                     & \hspace{-3mm}$\longrightarrow$ \ce{CH3+} + \ce{C3H5} + \ce{He} & 0.2 & 3.1$\times10^{-9}$ & 0.56 & 10 & 800 & 4 & \tablenotemark{w} & \rownumber \\
                     & \hspace{-3mm}$\longrightarrow$ \ce{C2H5+} + \ce{C2H3} + \ce{He} & 0.2 & 3.1$\times10^{-9}$ & 0.56 & 10 & 800 & 4 & \tablenotemark{w} & \rownumber \\
                     & \hspace{-3mm}$\longrightarrow$ \ce{C2H5} + \ce{C2H3+} + \ce{He} & 0.2 & 3.1$\times10^{-9}$ & 0.56 & 10 & 800 & 4 & \tablenotemark{w} & \rownumber \\
                     & \hspace{-3mm}$\longrightarrow$ \ce{CH2+} + \ce{CH3CHCH2} + \ce{He} & 0.2 & 3.1$\times10^{-9}$ & 0.56 & 10 & 800 & 4 & \tablenotemark{w} & \rownumber \\
\ce{C5H8} + \ce{H3+} & \hspace{-3mm}$\longrightarrow$ \ce{CH4} + \ce{C4H5+} + \ce{H2} & 0.2 & 4.2$\times10^{-9}$ & 0.65 & 10 & 800 & 4 & \tablenotemark{w} & \rownumber \\
                     & \hspace{-3mm}$\longrightarrow$ \ce{CH3+} + \ce{CH2CHCHCH2} + \ce{H2} & 0.2 & 4.2$\times10^{-9}$ & 0.65 & 10 & 800 & 4 & \tablenotemark{w} & \rownumber \\
                     & \hspace{-3mm}$\longrightarrow$ \ce{C2H5} + \ce{C3H4+} + \ce{H2} & 0.2 & 4.2$\times10^{-9}$ & 0.65 & 10 & 800 & 4 & \tablenotemark{w} & \rownumber \\
                     & \hspace{-3mm}$\longrightarrow$ \ce{C2H4+} + \ce{C3H5} + \ce{H2} & 0.2 & 4.2$\times10^{-9}$ & 0.65 & 10 & 800 & 4 & \tablenotemark{w} & \rownumber \\
                     & \hspace{-3mm}$\longrightarrow$ \ce{CH3CHCH2} + \ce{C2H3+} + \ce{H2} & 0.2 & 4.2$\times10^{-9}$ & 0.65 & 10 & 800 & 4 & \tablenotemark{w} & \rownumber \\
\ce{C5H8} + \ce{HCO+} & \hspace{-3mm}$\longrightarrow$ \ce{CH4} + \ce{C4H5+} + \ce{CO} & 0.2 & 1.4$\times10^{-9}$ & 0.65 & 10 & 800 & 4 & \tablenotemark{w} & \rownumber \\
                      & \hspace{-3mm}$\longrightarrow$ \ce{CH3+} + \ce{CH2CHCHCH2} + \ce{CO} & 0.2 & 1.4$\times10^{-9}$ & 0.65 & 10 & 800 & 4 & \tablenotemark{w} & \rownumber \\
                      & \hspace{-3mm}$\longrightarrow$ \ce{C2H5} + \ce{C3H4+} + \ce{CO} & 0.2 & 1.4$\times10^{-9}$ & 0.65 & 10 & 800 & 4 & \tablenotemark{w} & \rownumber \\
                      & \hspace{-3mm}$\longrightarrow$ \ce{C2H4+} + \ce{C3H5} + \ce{CO} & 0.2 & 1.4$\times10^{-9}$ & 0.65 & 10 & 800 & 4 & \tablenotemark{w} & \rownumber \\
                      & \hspace{-3mm}$\longrightarrow$ \ce{CH3CHCH2} + \ce{C2H3+} + \ce{CO} & 0.2 & 1.4$\times10^{-9}$ & 0.65 & 10 & 800 & 4 & \tablenotemark{w} & \rownumber \\
\ce{C5H8} + \ce{C+} & \hspace{-3mm}$\longrightarrow$ \ce{CH3} + \ce{C4H5+} + \ce{C} & 0.17 & 2.1$\times10^{-9}$ & 0.65 & 10 & 800 & 4 & \tablenotemark{w} & \rownumber \\
                    & \hspace{-3mm}$\longrightarrow$ \ce{C2H4} + \ce{C3H4+} + \ce{C} & 0.17 & 2.1$\times10^{-9}$ & 0.65 & 10 & 800 & 4 & \tablenotemark{w} & \rownumber \\
                    & \hspace{-3mm}$\longrightarrow$ \ce{C2H4+} + \ce{CH3CCH} + \ce{C} & 0.17 & 2.1$\times10^{-9}$ & 0.65 & 10 & 800 & 4 & \tablenotemark{w} & \rownumber \\
                    & \hspace{-3mm}$\longrightarrow$ \ce{C2H4+} + \ce{CH2CCH2} + \ce{C} & 0.17 & 2.1$\times10^{-9}$ & 0.65 & 10 & 800 & 4 & \tablenotemark{w} & \rownumber \\
                    & \hspace{-3mm}$\longrightarrow$ \ce{C3H5} + \ce{C2H3+} + \ce{C} & 0.17 & 2.1$\times10^{-9}$ & 0.65 & 10 & 800 & 4 & \tablenotemark{w} & \rownumber \\
                    & \hspace{-3mm}$\longrightarrow$ \ce{CH2+} + \ce{CH2CHCHCH2} + \ce{C} & 0.17 & 2.1$\times10^{-9}$ & 0.65 & 10 & 800 & 4 & \tablenotemark{w} & \rownumber \\
\ce{C5H8} + \ce{He+} & \hspace{-3mm}$\longrightarrow$ \ce{CH3} + \ce{C4H5+} + \ce{He} & 0.17 & 3.7$\times10^{-9}$ & 0.65 & 10 & 800 & 4 & \tablenotemark{w} & \rownumber \\
                     & \hspace{-3mm}$\longrightarrow$ \ce{C2H4} + \ce{C3H4+} + \ce{He} & 0.17 & 3.7$\times10^{-9}$ & 0.65 & 10 & 800 & 4 & \tablenotemark{w} & \rownumber \\
                     & \hspace{-3mm}$\longrightarrow$ \ce{C2H4+} + \ce{CH3CCH} + \ce{He} & 0.17 & 3.7$\times10^{-9}$ & 0.65 & 10 & 800 & 4 & \tablenotemark{w} & \rownumber \\
                     & \hspace{-3mm}$\longrightarrow$ \ce{C2H4+} + \ce{CH2CCH2} + \ce{He} & 0.17 & 3.7$\times10^{-9}$ & 0.65 & 10 & 800 & 4 & \tablenotemark{w} & \rownumber \\
                     & \hspace{-3mm}$\longrightarrow$ \ce{C3H5} + \ce{C2H3+} + \ce{He} & 0.17 & 3.7$\times10^{-9}$ & 0.65 & 10 & 800 & 4 & \tablenotemark{w} & \rownumber \\
                     & \hspace{-3mm}$\longrightarrow$ \ce{CH2+} + \ce{CH2CHCHCH2} + \ce{He} & 0.17 & 3.7$\times10^{-9}$ & 0.65 & 10 & 800 & 4 & \tablenotemark{w} & \rownumber \\
\ce{C6H8} + \ce{H3+} & \hspace{-3mm}$\longrightarrow$ \ce{CH4} + \ce{C5H5+} + \ce{H2} & 0.14 & 4.3$\times10^{-9}$ & 1.3 & 10 & 800 & 4 & \tablenotemark{w} & \rownumber \\
                     & \hspace{-3mm}$\longrightarrow$ \ce{C2H6} + \ce{C4H3+} + \ce{H2} & 0.14 & 4.3$\times10^{-9}$ & 1.3 & 10 & 800 & 4 & \tablenotemark{w} & \rownumber \\
                     & \hspace{-3mm}$\longrightarrow$ \ce{C2H5+} + \ce{CH2CHC2H} + \ce{H2} & 0.14 & 4.3$\times10^{-9}$ & 1.3 & 10 & 800 & 4 & \tablenotemark{w} & \rownumber \\
                     & \hspace{-3mm}$\longrightarrow$ \ce{C3H7} + \ce{l-C3H2+} + \ce{H2} & 0.14 & 4.3$\times10^{-9}$ & 1.3 & 10 & 800 & 4 & \tablenotemark{w} & \rownumber \\
                     & \hspace{-3mm}$\longrightarrow$ \ce{C2H2} + \ce{C4H7+} + \ce{H2} & 0.14 & 4.3$\times10^{-9}$ & 1.3 & 10 & 800 & 4 & \tablenotemark{w} & \rownumber \\
                     & \hspace{-3mm}$\longrightarrow$ \ce{C2H+} + \ce{C4H8} + \ce{H2} & 0.14 & 4.3$\times10^{-9}$ & 1.3 & 10 & 800 & 4 & \tablenotemark{w} & \rownumber \\
                     & \hspace{-3mm}$\longrightarrow$ \ce{C6H7+} + \ce{H2} + \ce{H2} & 0.14 & 4.3$\times10^{-9}$ & 1.3 & 10 & 800 & 4 & \tablenotemark{w} & \rownumber \\
\ce{C6H8} + \ce{HCO+} & \hspace{-3mm}$\longrightarrow$ \ce{CH4} + \ce{C5H5+} + \ce{CO} & 0.14 & 1.4$\times10^{-9}$ & 1.3 & 10 & 800 & 4 & \tablenotemark{w} & \rownumber \\
                      & \hspace{-3mm}$\longrightarrow$ \ce{C2H6} + \ce{C4H3+} + \ce{CO} & 0.14 & 1.4$\times10^{-9}$ & 1.3 & 10 & 800 & 4 & \tablenotemark{w} & \rownumber \\
                      & \hspace{-3mm}$\longrightarrow$ \ce{C2H5+} + \ce{CH2CHC2H} + \ce{CO} & 0.14 & 1.4$\times10^{-9}$ & 1.3 & 10 & 800 & 4 & \tablenotemark{w} & \rownumber \\
                      & \hspace{-3mm}$\longrightarrow$ \ce{C3H7} + \ce{l-C3H2+} + \ce{CO} & 0.14 & 1.4$\times10^{-9}$ & 1.3 & 10 & 800 & 4 & \tablenotemark{w} & \rownumber \\
                      & \hspace{-3mm}$\longrightarrow$ \ce{C2H2} + \ce{C4H7+} + \ce{CO} & 0.14 & 1.4$\times10^{-9}$ & 1.3 & 10 & 800 & 4 & \tablenotemark{w} & \rownumber \\
                      & \hspace{-3mm}$\longrightarrow$ \ce{C2H+} + \ce{C4H8} + \ce{CO} & 0.14 & 1.4$\times10^{-9}$ & 1.3 & 10 & 800 & 4 & \tablenotemark{w} & \rownumber \\
                      & \hspace{-3mm}$\longrightarrow$ \ce{C6H7+} + \ce{H2} + \ce{CO} & 0.14 & 1.4$\times10^{-9}$ & 1.3 & 10 & 800 & 4 & \tablenotemark{w} & \rownumber \\
\ce{C6H8} + \ce{C+} & \hspace{-3mm}$\longrightarrow$ \ce{CH3} + \ce{C5H5+} + \ce{C} & 0.2 & 2.2$\times10^{-9}$ & 1.3 & 10 & 800 & 4 & \tablenotemark{w} & \rownumber \\
                    & \hspace{-3mm}$\longrightarrow$ \ce{C2H5+} + \ce{C4H3} + \ce{C} & 0.2 & 2.2$\times10^{-9}$ & 1.3 & 10 & 800 & 4 & \tablenotemark{w} & \rownumber \\
                    & \hspace{-3mm}$\longrightarrow$ \ce{C2H5} + \ce{C4H3+} + \ce{C} & 0.2 & 2.2$\times10^{-9}$ & 1.3 & 10 & 800 & 4 & \tablenotemark{w} & \rownumber \\
                    & \hspace{-3mm}$\longrightarrow$ \ce{CH3CHCH2} + \ce{l-C3H2+} + \ce{C} & 0.2 & 2.2$\times10^{-9}$ & 1.3 & 10 & 800 & 4 & \tablenotemark{w} & \rownumber \\
                    & \hspace{-3mm}$\longrightarrow$ \ce{CCH} + \ce{C4H7+} + \ce{C} & 0.2 & 2.2$\times10^{-9}$ & 1.3 & 10 & 800 & 4 & \tablenotemark{w} & \rownumber \\
\ce{C6H8} + \ce{He+} & \hspace{-3mm}$\longrightarrow$ \ce{CH3} + \ce{C5H5+} + \ce{He} & 0.2 & 3.8$\times10^{-9}$ & 1.3 & 10 & 800 & 4 & \tablenotemark{w} & \rownumber \\
                     & \hspace{-3mm}$\longrightarrow$ \ce{C2H5+} + \ce{C4H3} + \ce{He} & 0.2 & 3.8$\times10^{-9}$ & 1.3 & 10 & 800 & 4 & \tablenotemark{w} & \rownumber \\
                     & \hspace{-3mm}$\longrightarrow$ \ce{C2H5} + \ce{C4H3+} + \ce{He} & 0.2 & 3.8$\times10^{-9}$ & 1.3 & 10 & 800 & 4 & \tablenotemark{w} & \rownumber \\
                     & \hspace{-3mm}$\longrightarrow$ \ce{CH3CHCH2} + \ce{l-C3H2+} + \ce{He} & 0.2 & 3.8$\times10^{-9}$ & 1.3 & 10 & 800 & 4 & \tablenotemark{w}  & \newtag{\rownumber}{rx:thisf}
    \label{tab:rxs}
    \enddata
\end{deluxetable*}
%\begin{minipage}[c]{\textwidth}
    %\footnotesize
	\tablecomments{Formulae of Type 1 and 2 are $k = \alpha \zeta$ and $k = \alpha e^{-\gamma A_\nu}$, where $k$ is in $\mathrm{s^{-1}}$, and formulae of Type 3 and 4 are $k(T) = \alpha \left(T/300\right)^\beta e^{-\gamma/T}$\\ and $k(T) = \alpha \beta \left(0.62 + 0.4767 \gamma \left(300/T\right)^{0.5}\right)$, where $k$ is in $\mathrm{cm^3\,s^{-1}}$ and $T$ is in $\mathrm{K}$, respectively.}
 \tablenotetext{a}{\citet{Doddipatla:2021}}
 \tablenotetext{b}{\citet{Majumdar:2013}}
 \tablenotetext{c}{\citet{Loison:2017} }
 \tablenotetext{d}{\citet{Hebrard:2013}}
 \tablenotetext{e}{\citet{tsang:1986}}
 \tablenotetext{f}{\citet{hoyermann:1979}}
 \tablenotetext{g}{\citet{Hebrard:2009}}
 \tablenotetext{h}{\citet{florescu-mitchell:2006}}
 \tablenotetext{i}{\cite{larsson:2005}}
 \tablenotetext{j}{\cite{janev:2004}}
 \tablenotetext{k}{\cite{reiter:2010}}
 \tablenotetext{l}{\citet{angelova:2004}}
 \tablenotetext{m}{\citet{tsang:1988}}
 \tablenotetext{n}{\cite{tsang:1991}}
 \tablenotetext{o}{\citet{Bohme:1982}}
 \tablenotetext{p}{\citet{anicich_index_2003}}
 \tablenotetext{q}{\citet{lias:1976}}
 \tablenotetext{r}{\citet{Baulch:1992}}
 \tablenotetext{s}{\cite{berteloite:2008}}
 \tablenotetext{t}{\citet{Morales:2010}}
 \tablenotetext{u}{\citet{canosa:1997}}
 \tablenotetext{v}{\cite{nizamov:2004}}
 \tablenotetext{w}{Estimated as part of this work (Sec.~\ref{sec:model})}

%\end{minipage}
\end{longrotatetable}

%\begin{comment}
%\begin{table}[]
%    \centering
%    \begin{tabular}{ c c c}
%    \toprule
%    	Species	&	$\mu_D$	&	$\alpha$\\
%                &   (D)     &  ($\AA^3$)
%    	\midrule
%    \ce{C9H8} & 0.73$^{\dagger}$ & 14.691$^{\ddagger}$ \\
%  \ce{C9H7CN} & 4.62$^{\ominus}$ & 17.1 \\
%    \ce{C4H8} & 0.438$^{\star}$  & 7.8 \\
%    \ce{C5H8} & 0.583$^{\odot}$  & 9.871 \\
%    \ce{C6H8} & 1.196$^{\dagger}$& 10.49 
%    \end{tabular}
%    \bottomrule
%    \caption{}
%    \label{tab:dipole_n_polar}
%\end{table}
%\end{comment}

%Dipole moment and polarizability used for calculation of rates of ion-neutral reactions (see Table~\ref{tab:rxs}).\\
%    $\dagger$ Calculated here at the $\omega$B97X-D/6-31+G(d) level of theory.\\
%    $\ddagger$ Calculated at $\omega$B97X-D/6-31+G** level of theory by NIST Computational Chemistry Comparison and Benchmark Database}\\
%    $\ominus$ \citet{Doddipatla:2021}\\
%    $\star$ \citet{S. Kondo, E. Hirota, and Y. Morino, J. Mol. Spectrosc. 28, 471 (1968)}\\
%    $\odot$ \citet{S.L. Hsu and W.H. Flygare, J. Chem. Phys. 52, 1053 (1970).

%TC:endignore

\end{document}